\documentclass[%
 reprint,
 amsmath,amssymb,
 aps,
prb,
]{revtex4-2}

\usepackage{natbib}
\usepackage{tikz}
\usepackage{graphicx}
\usepackage{dcolumn}
\usepackage{bm}

\usepackage[colorlinks=true, linkcolor=blue, citecolor=blue, urlcolor=blue]{hyperref}

\def\ldef{\mathrel{\mathop:}=}

\def\mysec#1{{\bf #1 --}}
\def\ev#1#2#3{\left\langle#1\right|#2\left|#3\right\rangle}
\def\inner#1#2{\left\langle#1\middle|#2\right\rangle}
\def\ket#1{\left|#1\right\rangle}

\def\G{{\mathbf{G}}}
\def\V{{\mathbf{V}}}
\def\E{{\mathbf{E}}}
\def\C{{\mathbf{C}}}

\def\dG{{\mathbf{G^*}}}
\def\dV{{\mathbf{V^*}}}
\def\dE{{\mathbf{E^*}}}

\def\dVf{{\mathbf{V_0^*}}}
\def\cl{{\mathcal{C}}}

\def\N{{|\V|}}
\def\M{{|\E|}}

\def\citeSM#1{ (see Appendix~\ref{sec#1})}

\tikzset{diagramnode/.style={fill=black, circle, inner sep=1pt}}

\newcommand{\vertex}{
  \tikz[baseline=-0.5ex]{\node[diagramnode] at (0,0) {};}
}

\newcommand{\tree}{
  \tikz[baseline=-0.5ex]{
    \node[diagramnode] (a) at (0,0) {};
    \node[diagramnode] (b) at (0.2,0) {};
    \node[diagramnode] (c) at (-0.2,0) {};
    \draw (a) -- (b);
    \draw (a) -- (c);
  }
}

\newcommand{\treeL}{
  \tikz[baseline=-0.5ex, yshift=-0.1em]{
    \node[diagramnode] (a) at (0,0) {};
    \node[diagramnode] (b) at (0.2,0) {};
    \node[diagramnode] (c) at (0.2,0.2) {};
    \draw (a) -- (b) -- (c);
  }
}

\newcommand{\loopSquare}{
  \tikz[baseline=-0.5ex, yshift=-0.1em]{
    \node[diagramnode] (a) at (0,0) {};
    \node[diagramnode] (b) at (0.2,0) {};
    \node[diagramnode] (c) at (0.2,0.2) {};
    \node[diagramnode] (d) at (0,0.2) {};
    \draw (a) -- (b) -- (c) -- (d) -- (a);
  }
}

\begin{document}

\title{Exact Ground States of Two Dimensional $\pm J$ Spin Glasses}

\author{Chaoming Song}%
\email{c.song@miami.edu}
\affiliation{%
Department of Physics, University of Miami, Coral Gables FL, 33146 USA.
}%

\date{\today}

\begin{abstract}
We derive exact analytical expressions for the ground-state energy and entropy of the two-dimensional $\pm J$ Ising spin glass, uncovering a nested hierarchy of frustrations. Each level in this hierarchy contributes through the kernel and pseudo-determinant of effective operators, capturing the energy and entropy, respectively. At leading order, the structure coincides with geometric plaquette frustrations, while subleading corrections arise from magnetic adjacency matrices defined on percolated clusters in the dual lattice. Our results, supported by numerical simulations, provide a systematic framework for analyzing spin-glass ground states and offer new insight into glass order in finite dimensions.
\end{abstract}

\maketitle

\mysec{Introduction}
Understanding spin-glass phases in finite-dimensional systems remains one of the deepest unsolved problems in statistical physics~\cite{mezard1987spin,nishimori2001statistical,talagrand2003spin,stein2013spin}. However, most existing results are numerical, and exact results are rare. Known analytical results predominantly exclude spin-glass ordering in two-dimensional models with continuous coupling distributions, instead demonstrating the uniqueness of the ground state~\cite{newman1992multiple,newman1996ground,newman1996spatial,newman1997metastate,newman2009metastates}. On the other hand, numerical studies strongly suggest that discrete $\pm J$ spin glasses possess a highly nontrivial and extensively degenerate ground-state structure~\cite{vannimenus1977theory,morgenstern1980magnetic,saul1993exact,blackman1998properties,kawashima1997finite,hartmann2000ground,hartmann2001lower,houdayer2001cluster,lukic2004critical,perez2012ground}. Especially in two dimensions, exact ground states for these models can be efficiently computed using polynomial-time graph algorithms, such as minimum-cut or matching algorithms~\cite{bieche1980ground,barahona1982computational,thomas2007matching}, or extrapolated from the low-temperature limit using Pfaffian methods~\cite{saul1993exact,galluccio2000new,lukic2004critical,jorg2006strong}. Thus, the two-dimensional $\pm J$ spin glass emerges as perhaps the simplest finite-dimensional candidate for genuine spin-glass ordering, although analytic results remain elusive.

In this Letter, we propose an explicit analytic formula for the exact ground-state energy and entropy of the two-dimensional $\pm J$ spin glass. Our results are expressed through a nested hierarchy of frustrations, characterized by a sequence of effective operators with nested kernels. The dimension of these kernels corresponds to the ground-state energy, and the associated pseudo-determinant contributes to the entropy. At the zeroth level, the frustration structure aligns with regular geometric plaquette frustrations~\cite{toulouse1987theory,barahona1982computational}. The level-$1$ frustration is governed by a magnetic adjacency matrix associated with percolated clusters in the dual lattice~\cite{stein1987ground,sunada1994discrete,machta2008percolation}. Higher levels involve percolation beyond nearest neighbors. We find that at the first level, a novel frustration circulation law emerges, under which the ground state is entirely determined by the percolation geometry. We test our result numerically and find excellent agreement with previous results. Our framework not only enables analytic computation of the zero-temperature entropy of the 2D $\pm J$ spin glass, but also opens a new direction for analytic understanding of spin-glass order in finite dimensions.

\mysec{Self-dual formula}
We consider the Ising spin glass~\cite{edwards1975theory} defined on a planar graph $\G = (\V, \E)$, with the Hamiltonian
\begin{equation}
    H = -\sum_{(ij)\in \E} J_{ij} \sigma_i\sigma_j,  
    \notag
\end{equation}
where the coupling constant $J_e$ is assigned to each edge $e \in \E$. Our discussion primarily focuses on $\pm J$ disorder on a square lattice, i.e., $P(J_e) = (1-p)\delta(J_e - 1) + p\, \delta(J_e + 1)$.

\begin{figure}
  \includegraphics[width=1\linewidth]{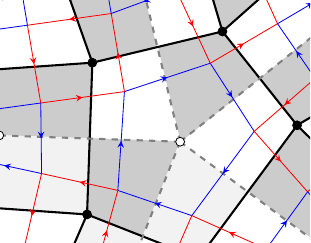}
  \caption{Embedding of $\G$ and its dual $\dG$. Curl operators $D$ (red) and ${D^*}^\dag$ (blue) act around vertices (black) and dual vertices (white).}
  \label{fig:graph}
\end{figure}

We build on a recently discovered combinatorial self-dual formula for the partition function~\cite{song2024kramers}. Assuming an isoradial embedding of the dual graph $\dG = (\dV, \dE)$ over $\G$, with mutually dual vertices and faces and orthogonal edge pairs (Fig.~\ref{fig:graph}), we obtain a tiling of $2|\E|$ quadrilaterals. Each quadrilateral links a primal-dual vertex pair $(v, v^*)$ via adjacent edges in $\G$ and $\dG$. Using this structure, we derive\begin{align}
    Z^2 = 2^{\N}\cosh(2\beta J)^\M\det\left(I - U\right),
\label{eq:mine1}
\end{align}
where the $2|\E| \times 2|\E|$ matrix
\begin{equation}
    U \ldef \frac{2x}{1 + x^2} D + \frac{1 - x^2}{1 + x^2} {D^*}^\dag,
    \notag
\end{equation}
with $x \ldef e^{-2\beta}$. Here, $D = \sum_{v \in \V} D_v$ and $D^* = \sum_{v^* \in \dV} D_{v^*}$ act on the quadrilateral basis, representing local curls around primal and dual vertices and serving as order and disorder operators~\cite{kadanoff1971determination,fradkin2017disorder}. Their matrix elements
\begin{align} \ev{q}{D_v}{q'} = e^{i \gamma(q, q')/2}, \; \ev{q}{D_{v^*}}{q'} = J_e e^{i \gamma^*(q, q')/2},
\label{eq:weight}
\end{align} are nonzero when $q$ and $q'$ share an edge $e$, with $q'$ counterclockwise from $q$. The angles $\gamma, \gamma^*$ track rotations around $v$ and $v^*$ (Fig.~\ref{fig:graph}).

The operator $U$ is self-dual under $G \leftrightarrow G^*$ and $x \leftrightarrow x^* = (1-x)/(1+x)$, manifesting Kramers--Wannier duality. Moreover, $U$ is unitary, implying a circle law for $Z$~\cite{song2025circle}.

\mysec{Ground State Hierarchy}
We now turn our attention to the zero-temperature limit $\beta \to \infty$ of Eq.~\eqref{eq:mine1}. In this limit, $x$ becomes exponentially small, so $U = \sum_{v^* }{D_{v^*}}^\dag + O(x)$ is dominated by disorder operators. It is block-diagonalized into $|\dV|$ blocks, each associated with a dual vertex $v^*$. For each block, a direct computation yields
\begin{align}
    \det(I-D_{v^*}^\dag) = 1+W_{v^*},
\notag
\end{align}
where $W_{v^*} \equiv \prod_{e\in E(v^*)}J_e = \pm 1$ is defined as the product of edge disorders around the dual vertex $v^*$. This is known as the geometrical frustration, which we refer to as level-0 frustration; its interpretation will become clear later. When the plaquette is frustrated, i.e., $W_{v^*} = -1$, it contributes a $\Phi_0 = \pi$ flux, resulting in destructive interference. As a result, the corresponding determinant vanishes and acquires a correction from the next order.

To capture these corrections, note that for each frustrated dual vertex, there exists a unique null vector $\ket{v^*}$ satisfying
\begin{equation}
    D_{v^*}^\dag \ket{v^*} = \ket{v^*},
\label{eq:eigen}
\end{equation}
where the component $\inner{q}{v^*} = \frac{1}{\sqrt{d_{v^*}}}e^{i\theta(q,v^*)}$ if the quadrilateral $q$ is connected to $v^*$, with some phase $\theta$ determined by the disorder bond $J_e$ attached to $v^*$. Denoting $V_0^* \subset V^*$ as the subset of frustrated dual vertices, the kernal $\ker(I-D^\dag)$ is spanned by the vectors $\ket{v^*}$ for $v^* \in V_0^*$. Consequently, the determinant decomposes via Schur complements into the kernel and image of the operator $D^*$ as
\begin{align}
\ln\det &(I-U) = \dim\ker(I-{D^*}^\dag) \ln x+\ln{\det}'(I-{D^*}^\dag)\notag \\ 
& +\ln\det\left(-D+O(x)|_{\ker(I-{D^*})}\right)  + O(x),
\notag
\end{align}
where ${\det}'$ denotes the pseudo-determinant, i.e., the product of non-zero eigenvalues. This can be computed directly as ${\det}'(I-{D^*}^\dag) = (|\dV|-|\dVf|)\ln 2 + \sum_{v^* \in \dVf} \ln d_{v^*}$.

Similarly, we need to examine the kernel of $i H_1 \ldef -D|_{\ker(I-{D^*})}$ at the next level. By proceeding recursively, we obtain a hierarchy of frustrations and their resolution via Schur complements\citeSM{1}:
\begin{align}
&E_0 = -\M + |\dVf|+\sum_i \dim \ker(H_i), \notag\\ 
&S_0 = \frac{1}{2}  \left(\sum_{v^* \in \dVf} \ln d_{v^*} +\sum_i \ln \det{}'(iH_i) +\chi_E\ln2 \right),
\notag
\end{align}
where $\chi_E \ldef |V|-|E| + |V^*|$ is the Euler characteristic. The operators $H_i$ are nested at each level, each acting on the kernel from the previous level, generating a chain complex $\ker(H_1) \supset\ker(H_2) \supset \ldots$. This allows us to compute the ground state recursively level by level.

At level zero, note that for a dual vertex $v^*$ with degree $d_{v^*}$, the probability of having $W_{v^*} = -1$ is given by $f(d_{v^*}) \ldef \frac{1-(2p-1)^{d_{v^*}}}{2}$. In other words, the expected dimension of the kernel is $\overline{|\dVf|} = \sum_{v^* \in \dV} f(d_{v^*})$, and $\overline{\sum_{v^* \in \dVf} \ln d_{v^*} } = \sum_{v^* \in \dV} f(d_{v^*})\ln d_{v^*}$, where the overline denotes the quenched average. These expressions provide a lower bound on the ground state energy, consistent with results obtained using geometric approaches~\cite{vannimenus1977theory,grinstein1979ising}. Additionally, they also provide an upper bound on the ground state entropy.

For higher levels, we performed numerical simulations on a $100\times100$ square lattice, averaged over $10^4$ realizations\citeSM{4}. At level 1, we find $e_0 \approx -1.418$ and $s_0 \approx 0.102$, while at level 2, we obtain $e_0 \approx -1.407$ and $s_0 \approx 0.066$. These results are consistent with existing numerical estimates $e_0 \approx -1.402$ \cite{zhan2000new,lukic2004critical,kobe2006ground} and $s_0 \approx 0.071$ \cite{hartmann2000ground,blackman1998properties,lukic2004critical,kobe2006ground}. The slightly lower entropy may be due to the fact that there is no direct numerical estimate of the entropy; instead, previous works rely on extrapolation from low temperature or Monte Carlo simulations, which may slightly overestimate the result.

\begin{figure}[htbp]
  \includegraphics[width=1\linewidth]{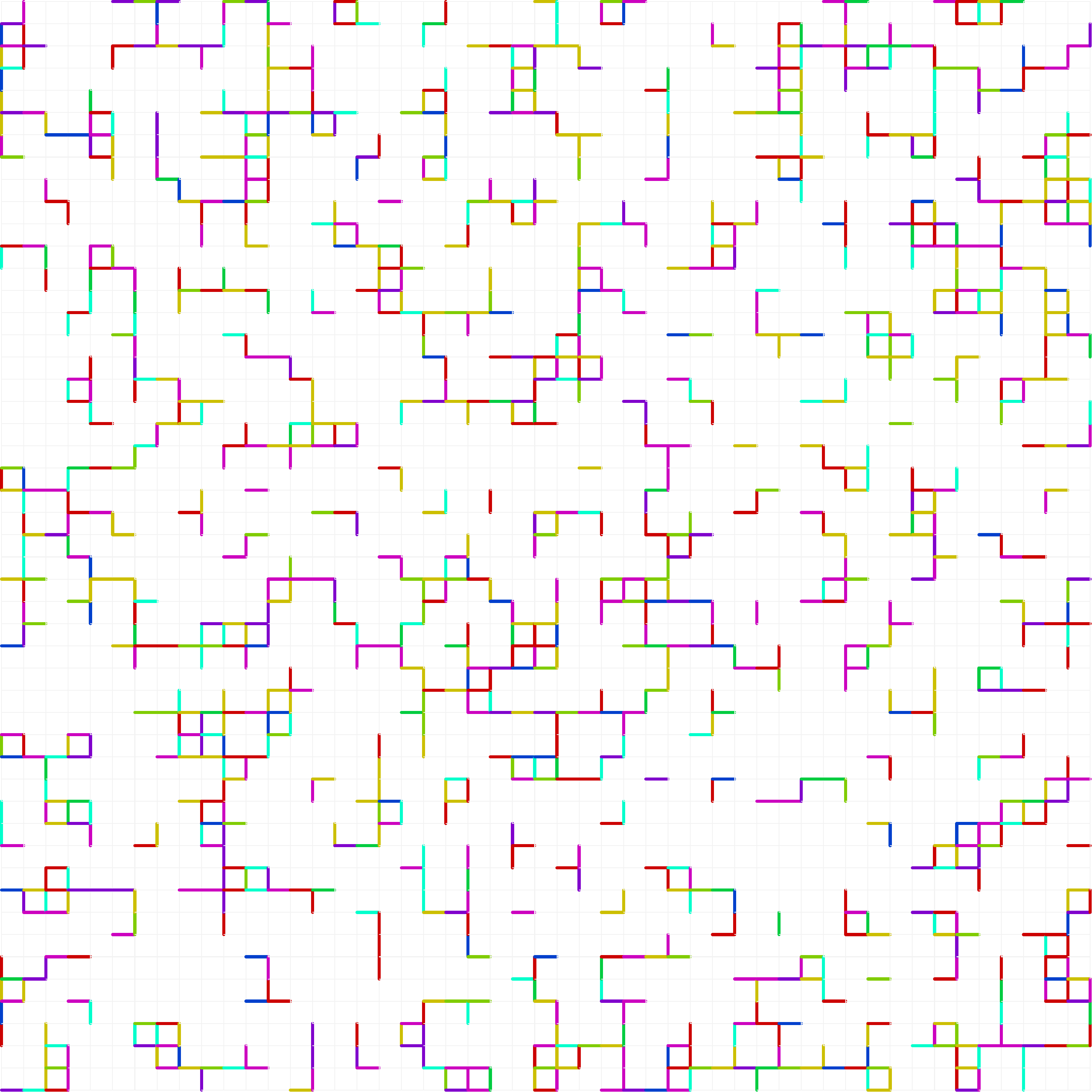}
  \caption{The magnetic adjacency matrix $H_1$ on the square lattice, restricted to percolated clusters. Bond colors represent the Peierls phase $\phi$.}
  \label{fig:perc}
\end{figure}

\mysec{Level-1 Frustration}
We now examine in detail the ground state energy and entropy at level-1 frustration. The corresponding operator is defined as \( H_1 \ldef iD|_{\ker(I - {D^*}^\dag)} \). As it is restricted to the kernel of \( I - {D^*}^\dag \), which we discussed previously, this space corresponds to the vector space spanned by percolated clusters in the dual lattice, where each dual site \( v^* \) is occupied with probability \( f(d_{v^*}) \) (see Figure~\ref{fig:perc}). Note that unless \( p = 1/2 \), i.e., \( f(d_{v^*}) = 1/2 \), the occupation is correlated due to the projection from \( J \) to \( W \).

Equation~\eqref{eq:weight} shows that \( H_1 \) is a normalized magnetic adjacency matrix
\begin{equation}
    \ev{{v^*}'}{H_1}{v^*} = \frac{1}{\sqrt{d_{{v^*}'}d_{v^*}}} A^*_{{v^*}',v^*} e^{i\phi_{{v^*}',v^*}},
\notag
\end{equation}
where \( A^* \) is the ordinary adjacency matrix on the dual lattice, restricted to the percolated clusters. The Peierls phase \( \phi \) is skew-symmetric and incorporates the \( U(1) \) holonomy \cite{peierls1933theorie}, making \( H_1 \) Hermitian\citeSM{2}. This follows from Eqs.~(\ref{eq:weight}--\ref{eq:eigen}).

The phase \( \phi \) is gauge-dependent, meaning it can be altered by a local phase shift at each site
\(
\ket{v^*} \rightarrow e^{i \theta_{v^*}}\ket{v^*}.
\)
Under such a gauge transformation, the Peierls phase transforms as
\(
\phi_{ij} \rightarrow \phi_{ij} + \theta_i - \theta_j.
\)
However, the magnetic flux through a fundamental cycle is gauge-invariant. Since each fundamental cycle of \( \dG \) corresponds to a vertex \( v \in \V \), we denote them as \( C_v \). The corresponding magnetic flux around \( C_v \) is defined as
\(
\Phi_1(C_v) \ldef \sum_{(i,j) \in C_v} \phi_{ij},
\)
modulo \( 2\pi \). It can be shown that
\begin{equation}
    \Phi_1(C_v) = (1 + d_v/2)\pi \mod 2\pi,
\notag
\end{equation}
where the \( \pi \) phase arises from the overall phase of the curl operator \( D_v \) around the cycle \( C_v \), and the \( d_v \pi/2 \) phase results from the \( d_v \) multiplicity of the imaginary unit in the definition of \( H_1 \).

More generally, the phase accumulated around any simple cycle $C$ is quantized according to the enclosed frustration, obeying the relation\citeSM{3}
\begin{equation}
   \sum_{(i,j) \in C} \left(\phi_{ij}-\frac{\pi}{2}\right) = \left(1+n_f(C)\right)\pi \mod 2\pi,
\label{eq:circulation}
\end{equation}
where $n_f(C)$ is the number of frustrated dual vertices enclosed by $C$. This {\it Frustration Circulation Law} expresses a discrete topological constraint: frustration acts as a quantized source of phase accumulation, fixing the holonomy around any loop in terms of enclosed defect number. In this sense, frustration generates nontrivial lattice circulation in the absence of an external gauge field.

Since the magnetic adjacency matrix \( H_1 \) acts on percolated sites, it is natural to decompose \( H_1 = \bigoplus_\cl A^*_\phi(\cl) \) into blocks over percolated clusters. Here \( \cl \) denotes isolated percolated clusters, and \( A^*_\phi(\cl) \) is the corresponding magnetic adjacency matrix. We distinguish \( \cl \) based on their embedding in the dual lattice. In addition, for non-simply connected clusters, as suggested by the circulation law, Eq.~\eqref{eq:circulation}, we also distinguish \( \cl \) based on the odd or even number of frustrations enclosed by every loop, which play an important role, as we will see later. Accordingly, we have
\(
\dim \ker H_1 = \sum_\cl \dim \ker A^*_\phi(\cl)
\) and  $\ln {\det}'(iH_1) = \sum_{\cl} \ln {\det}'(iA_\phi^*(\cl))$.

To calculate the kernel and pseudodeterminant, a natural way to compute \( {\det}'(iH_1) \) is by calculating the corresponding nonzero eigenvalues. These appear in \( \pm \lambda \) pairs, ensuring that the entropy is real. Moreover, the spectral radius is less than one, i.e., the eigenvalues lie in \( (-1, 1) \), implying \( \ln {\det}'(iH_1) < 0 \). However, in many cases, such as the square lattice, the site occupation probability is below the percolation threshold $f_c \approx 0.592746$, and there is no giant cluster. Consequently, the spectrum of \( H_1 \) remains non-compact yet densely discrete in the thermodynamic limit. This arises because each small cluster contributes its own discrete spectral component, resulting in a dense but non-continuous spectrum near zero. Therefore, studying the spectral density is not suitable in the thermodynamic limit.

Instead, we consider the characteristic polynomial of \( -i A^*_\phi(\cl) \), given by \( p(\lambda) =\det(\lambda I + i A^*_\phi(\cl)) = \lambda^n + c_{1} \lambda^{n-1} + \ldots + c_n \), where \( n = |\cl| \) is the size of \( \cl \). Each coefficient \( c_k \) can be computed as a weighted sum over cycle covers on \( k \) vertices, 
\(
c_{k} = \sum_{|\C| = k} (-1)^{n(\C)} \prod_{(i\to j)\in \C} (-i A^*_\phi(\cl)_{ij}).
\)
Here, the cycle cover \( \C = \bigcup C_i \) is a disjoint union of oriented cycles \( C_i \), that is, a set of dimers (2-cycles) and regular cycles, each counted twice due to orientation. By incorporating the frustration circulation law \eqref{eq:circulation}, we obtain
\begin{equation}
    c_{k} =  \sum_{|\C| = k} (-1)^{n_f(\C)} \prod_{v^*\in \C} \frac{1}{d_{v^*}},
\end{equation}
where \( n_f(\C) \ldef \sum_i n_f(C_i) \) denotes the total frustration enclosed by the cycle cover \( \C \). The kernel size is given by \( \dim \ker A = n - r(\cl) \), where the rank \( r(\cl) \) is the largest integer for which the corresponding coefficient is nonzero. The pseudo-determinant \( \det' i A^*_\phi(\cl) = c_{r}(\cl) \) is then given by this coefficient. This result shows that the disorder-averaged ground state at level-1 is purely geometric; it depends only on the structure of the percolation clusters, not on the magnetic phase \( \phi \). This implies
\begin{align}
   \overline{\dim \ker H_1} &= \sum_{\cl} \left( |\cl| - r(\cl) \right) N(\cl), \notag\\ 
   \overline{\ln {\det}'(iH_1)} &= \sum_{\cl} \ln(c_r(\cl)) N(\cl),
   \notag
\end{align}
where \( N(\cl) \) is the expected number of clusters \( \cl \). For example, at \( p = 1/2 \), and for simply connected clusters, we have \( N(\cl) = 2^{-|\cl|-|\partial \cl|} |\dV|\), where \( \partial \cl \) denotes the boundary, that is, the set of neighboring vertices, which depends on the embedding. For clusters that contain holes, \( N(C) \) must be computed separately based on the frustration enclosed. When there is no giant cluster, such as in the square lattice, the number of clusters decays asymptotically with size. The above equation can then be expanded diagrammatically using small clusters.

Moreover, there is a connection between the rank \( r(\cl) \) and the maximum matching \( \nu(\cl) \) of a percolated cluster \( \cl \). Specifically, for an isolated cluster without odd-length cycles, as in the square lattice, we have the bound
\[
r(\cl) \leq 2\nu(\cl),
\]
where \( \nu(\cl) \) denotes the size of a maximum matching. This bound is saturated when \( \cl \) is a tree, since in the absence of cycles, the only non-vanishing cycle covers correspond to disjoint dimers. In this case, the magnetic phases \( \phi \) can be fully gauged away, and the rank reduces to that of the underlying tree graph, as shown in \cite{cvetkovic1972algebraic}. When cycles are present, the situation becomes more subtle. Even-length cycles admit two distinct perfect matchings, which contribute with opposite signs in the alternating sum defining the determinant. If these contributions cancel exactly, the coefficient \( c_{2\nu} \) vanishes, and the rank drops below the upper bound. However, so long as the alternating sum does not vanish due to contributions from cycles with odd frustration \( n_f(\C) \), the leading coefficient \( c_{2\nu} \) remains nonzero, and the bound is saturated. Therefore, the strict inequality \( r(\cl) < 2\nu(\cl) \) occurs only when all contributing cycles carry odd frustration.

To show the cluster expansion more explicitly, we focus on regular lattices with \( d_v = d \) and \( d_{v^*} = d^* \). The primary example is the square lattice, where \( d = d^* = 4 \). In this case, the ground state energy density \( \overline{e_0}^{(1)} \ldef \frac{1}{\N} \overline{\dim \ker H_1} \) is given by
\begin{equation}
   \overline{e_0}^{(1)} = P(\vertex)+2P(\tree) + 4P(\treeL)+ 4P(\loopSquare)\ldots,
\notag
\end{equation}
where the leading terms arise from the isolated vertex \and the 3-vertex path graph. At \( p = 1/2 \), the probabilities are given by \( P(\vertex) = 2^{-5} \), \( P(\tree) = 2^{-10} \), \( P(\treeL) = 2^{-11} \), and \( P(\loopSquare) = 2^{-12} \). The first term gives a lower bound \( \overline{e_0}/J > -47/32 = -1.46875 \), which accounts for approximately 38\% of the total and already improves upon existing bounds \cite{vannimenus1977theory,grinstein1979ising}.
Similarly, the level-1 entropy can be computed from the corresponding coefficients
\begin{equation}
    \ln c_{r}(\cl) = \ln \mathsf{m}(\cl) - r(\cl)\ln d^*,
\label{eq:det1}
\end{equation}
where \( \mathsf{m}(\cl) \ldef \sum_{|\C| = r} (-1)^{n_f(\C)} \) is a positive integer. The ground state entropy density is then
\begin{equation}
    \overline{s_0} = \frac{1}{2} \left( \overline{e_0}^{(1)} \ln d^* + \overline{\ln\mathsf{m}} \right) + \text{higher levels},
\notag
\end{equation}
where \( \overline{\ln\mathsf{m}} \ldef \frac{1}{\N}\sum_{\cl} \ln \overline{\mathsf{m}}(\cl) N(\cl) \). As with energy, \( \overline{\ln\mathsf{m}} \) can be expanded in cluster orders. For the square lattice,
\begin{equation}
\overline{\ln\mathsf{m}} = 2\ln 2 P(\tree) + 4\ln 2 P(\treeL)+ \ln 3 P(\loopSquare)+\ldots.
\notag
\end{equation}
Adding more clusters refines the bound, although saturation only occurs for cluster sizes in the range \( 200 \sim 300 \)\citeSM{4}.

\begin{figure}
\centering
  \includegraphics[width=0.8\linewidth]{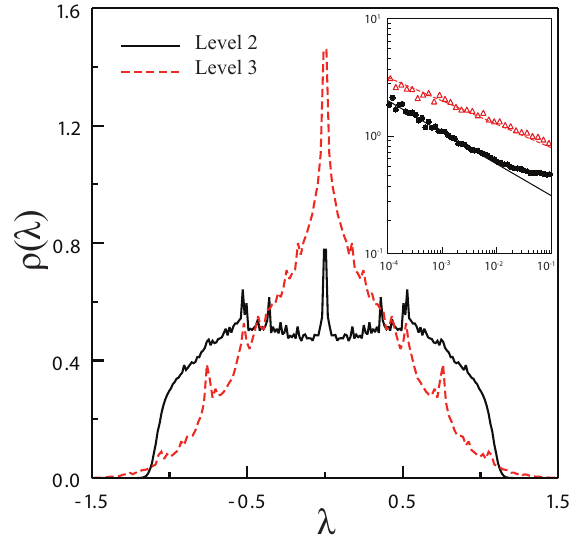}
  \caption{Spectral density $\rho(\lambda)$ for level-2 (black solid) and level-3 (red dashed) operators restricted to their respective giant clusters. Side peaks may reflect finite-size effects. Inset: log-log plot near $\lambda=0$ with fitted power-law guides (slopes $\alpha_2 \approx 2.7$, $\alpha_3 \approx 2.0$).}
  \label{fig:spectrum}
\end{figure}
\mysec{Higher Levels} 
We now consider higher levels, extending beyond nearest neighbours. For instance, $H_2 = i\left(2D(I-{D^*}^\dag)^+D-I\middle)\right|_{\ker{H_1}}$, where $+$ denotes the Moore--Penrose pseudoinverse. This operator involves next-nearest neighbour terms. Similarly, $H_3$ incorporates next-next-nearest neighbour interactions, and so forth. These operators remain block-decomposable into a direct sum over corresponding clusters $\cl'$, $\cl''$, $\ldots$. A level-2 cluster typically results from the fusion of several level-1 clusters projected to the kernel of $H_1$, corresponding to combinations of uncovered isolated vertices. As these vertices are typically long ranged, it then generates the long-range correlation. The properties of $H_i$ on higher-level clusters are less understood. In particular, whether a geometric characterization of $\ker H_i$ exists, analogous to level-1, remains unknown. On the other hand, preliminary numerical evidence suggests that ${\det}'(iH_2(\cl))$ is rational. Moreover, in analogy with Eq.~\eqref{eq:det1}, we observe the empirical relation
\begin{equation}
\ln {\det}'(iH_2(\cl')) = \ln \mathsf{m}_2(\cl') -\sum_{\cl \in \cl'}\mathsf{m}(\cl) -r_2(\cl')\ln d^*,
\notag
\end{equation}
where $\mathsf{m}_2(\cl')$ and $r_2(\cl')$ are integers, with $r_2(\cl') \leq \mathrm{rank}(iH_2(\cl'))$. Notably, the negative contribution to the level-1 entropy appears partially cancelled by higher levels, resulting in the residual bound
\(
(\dim \ker H_1 - \sum_{\cl'} r_2(\cl')) \ln d^* \geq \dim \ker H_2 \ln d^*.
\)
If the inequality is strict, and similar patterns hold at all higher levels, it would imply the existence of nonzero residual entropy, $S_0 > 0$.

As interaction range increases with level, the associated percolation thresholds decrease. Consequently, giant clusters appear for $H_i$ at high levels. This suggests that the spectrum of $H_i$ on such clusters may exhibit continuous components. Figure~\ref{fig:spectrum} shows the spectra $\rho_2(\lambda)$ and $\rho_3(\lambda)$ of $H_2$ and $H_3$ on giant clusters obtained from numerical simulations. The densities follow power-law scaling $\rho_i(\lambda)\sim \lambda^{-\alpha_i}$ near $\lambda = 0$, with exponents $0<\alpha_i<1$ indicating non-analytic behavior.

\mysec{Discussion}
Our results suggest that the ground state of the 2D $\pm J$ model is governed by a hierarchy of frustrations. We have analyzed the level-1 structure in detail, showing that it is determined entirely by the geometry of percolation clusters. Although the idea of using percolation to understand the $\pm J$ ground state has been proposed previously~\cite{stein1987ground,sunada1994discrete,machta2008percolation}, an exact and explicit correspondence, rather than a heuristic argument, has remained absent from the literature. Remarkably, we find that our construction is also closely related to the graph matching problem, reminiscent of exact numerical algorithms for computing ground state energies via the minimum-weight perfect matching (MWPM) framework~\cite{bieche1980ground, barahona1982computational,thomas2007matching,hartmann2011ground}. In this sense, our method systematically unfolds the MWPM formulation into a hierarchical structure, enabling analytical examination at each level. In practice, MWPM implementations require a distance cutoff to remain computationally feasible on large lattices\cite{barahona1982computational,bieche1980ground,thomas2007matching}. This is directly analogous to our level truncations.

In contrast to MWPM, which must be solved numerically, our approach allows for analytical computation of the quenched average ground state energy and entropy, at least at level-1. More significantly, it reveals an underlying algebraic and geometric structure in the ground state that is not accessible through conventional formulations. In particular, the level-wise decomposition highlights how ground state properties are organized according to the topology of frustration~\cite{toulouse1987theory}. Furthermore, our method yields an exact algorithm for both energy and entropy, with the latter being especially difficult to compute directly. On the square lattice, where giant clusters appear only at higher levels, the algorithm remains efficient due to the rapid decay of the kernel dimension with increasing level.

Several open questions remain. While we have rigorously established Hermiticity for the level-1 operator, a general proof for higher levels is still lacking. Numerical results at levels 2 and 3 support the Hermitian conjecture, but a complete analytic understanding remains an open challenge. Another important question is whether the geometric structure identified at level-1 generalizes to higher levels. Establishing such a structure would provide a constructive framework for determining the presence or absence of residual entropy in the ground state, which is a defining feature of a genuine glassy phase.  A related question is whether these methods can be adapted to three dimensions. Although the 3D case is not integrable and a simple determinant formula is unlikely, many of the geometric features found in two dimensions, such as the connection to percolation, may still carry over. Exploring this possibility could shed light on the nature of frustration and disorder in higher-dimensional systems. 

Moreover, the emergence of giant clusters at higher levels also leads to non-integer exponents in the spectral density near zero, with significant implications for low-temperature excitations. In conventional scenarios, excitations scale as integer powers $x^n$, where even $n$ corresponds to local spin-flip processes and odd $n$ to domain-wall rearrangements~\cite{lukic2004critical, jorg2006strong}. In contrast, the presence of giant clusters implies non-integer $n$, associated with fractional excitations arising from collective modes that are not localized to finite-size defects. Since giant clusters in the square lattice appear only at higher levels, such fractional modes are expected to be suppressed. It is therefore important to study other lattices, such as the hexagonal lattice, whose dual, the triangular lattice, has a percolation threshold well below $1/2$. In such systems, giant clusters, and hence fractional excitations, may already emerge at level-1. These extended modes, tied to long-range frustration structures, could leave observable signatures in experiments.

Finally, our approach is also applicable to systems with continuous disorder, and the corresponding results will be reported in detail elsewhere. Altogether, our theory introduces a new analytical framework for investigating zero-temperature glassy behavior in two-dimensional disordered systems and provides a systematic method for uncovering the hierarchical organization of their ground state structure.


\appendix

\section{Low-Temperature Expansion via Schur Complement}\label{sec1}
In this section, we provide a derivation of the formula in the main text. We start with operators $A_{n}$ and $B_{n}(x)$, where the latter depends on a variable $x$. We may assume $B_{n}(x)$ is analytic in $x$. The operators $A_{n}$ have a nontrivial kernel. By separating the kernel and image of the operators $A_{n}$, we have the following Schur complement decomposition:
\begin{align}
    \ln&\det(A_{n} + x B_{n}(x)) = \ln \det\left(A_{n}^{11}+xB_{n}^{11}(x)\right) \notag \\
    &+\ln\det\left(x B_{n}^{00}(x)- x^2 B_{n}^{01} \left(A_{n}^{11}+xB_{n}^{11}(x)\right)^{-1} B_{n}^{10} \right), \notag 
\label{eq:recur}
\end{align}
where superscripts $0$ and $1$ denote the projections onto the kernel and image, respectively. Equivalently, we obtain a recursive relation:
\begin{align}
    \ln&\det(A_{n} + x B_{n}(x)) =\ln\det(A_{n+1}+ x B_{n+1}(x)) \notag\\ 
    &+ \dim \ker A_{n} \ln x +  \ln {\det}'A_{n}  + O(x),
\end{align}
where $A_{n+1}$ and $B_{n+1}(x)$ are defined as
\begin{align}
    &A_{n+1} \ldef B_{n}(0)|_{\ker A_{n}},\notag\\
    &B_{n+1}(x) \ldef \left.\frac{B_{n}(x)-B_{n}(0)}{x} \right|_{\ker A_{n}} \notag\\
    &\quad - \left.
    B_{n}(x) \left(A_{n}+xB_{n}(x)\right)^{+} B_{n}(x)
    \right|_{\ker A_{n}}.\notag
\end{align}
Here, the restricted pseudo-inverse $+$ is defined as
\begin{equation}
    \left(A_{n}+xB_{n}(x)\right)^{+} \ldef A_n^{+} - x  A_n^{+} B_{n}(x) A_n^{+} + \ldots, \notag
\end{equation}
where $A_n^{+}$ denotes the regular Moore--Penrose pseudo-inverse of $A_n$.

In our setup, the zeroth order is given by
\[
A_0 = I - {D^*}^\dag,\; B_0(x)= - 2D + 2x (I - A^{(0)}).
\]
For technical convenience, we rescale the first order by a factor of two as
\(
A_1 = 2 \tilde A_1 ,\; B_1(x)= 2 \tilde B_1(x),
\)
leading to
\begin{align}
    &\tilde A_1 = -D|_{\ker A_0},\notag \\
    &\tilde B_1(x)= I - D(\tilde A_0^+ + x \tilde A_0^+ (D - x (I - A_0))\tilde A_0^+ + \ldots)D|_{\ker A_0}, \notag
\end{align}
where $\tilde A_0 \ldef A_0/2$. By doing this, we introduce an extra term $\dim \ker A_0 (\ln 2)$. Defining $iH_n \ldef \tilde A_n$ for $n \geq 1$, we obtain
\begin{align}
    H_1 = iD|_{\ker(I - {D^*}^\dag)},\quad H_2 = i (D \tilde A_0^+ D - I)|_{\ker(H_1)}, \ldots \notag
\end{align}
Applying Eq.~\eqref{eq:recur} recursively, we find
\begin{align}
    \ln\det&(A_0 + xB_0(x)) =  \sum_{i} \left( \dim\ker H_i \ln x + \ln {\det}'(iH_i) \right) \notag \\ 
    &+ \dim \ker A_0 \ln 2x +  \ln {\det}'A_0 + O(x), \notag
\end{align}
where the additional factor of two arises from the rescaling.

Equation~\eqref{eq:mine1} gives the partition function at small $x \ldef \exp(-2\beta)$ as
\begin{align}
    \ln Z = \frac{1}{2}\left(\N \ln 2 - \M \ln 2x + \ln\det( A_0+ x B_0(x))\right)  + O(x^2). \notag
\end{align}
Matching the low-temperature expansion
\begin{align}
    \ln Z = \frac{1}{2} E_0 \ln x + S_0 + O(x), \notag
\end{align}
we obtain
\begin{align}
    &E_0 = -\M + \dim \ker A_0 +  \sum_{i} \dim\ker H_i, \notag\\
    &S_0 = \frac{1}{2} ((\N - \M +  \dim \ker A_0) \ln 2 + \ln {\det}'A_0 \notag \\
    &\quad + \sum_i \ln {\det}'(iH_i) ). \notag
\end{align}
Using the fact that $\dim \ker A_0 = |\dVf|$ and ${\det}'A_0 = (|\dV| - |\dVf|)\ln 2 + \sum_{v^* \in \dVf} \ln d_{v^*}$, we obtain the formula in the main text.

\section{Proof of Hermitianity of $H_1$}\label{sec2}

\begin{center}
\refstepcounter{figure}
\includegraphics[width=0.5\linewidth]{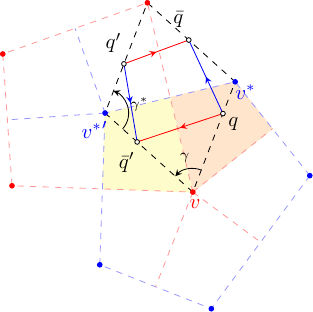}
\\[0.5em]
\small \textbf{Figure~\thefigure.} Illustration of two neighboring vertices and related quadrilaterals.
\label{fig:adj}
\end{center}

In this section, we prove that $H_1 = iD|_{\ker(I - {D^*}^\dag)}$ is Hermitian, that is, $D$ is anti-Hermitian when restricted to the kernel of $I - {D^*}^\dag$. Since we are working within the kernel, for any $v^* \in \dVf$, it satisfies
\begin{align}
{D^*}^\dag |v^*\rangle = |v^*\rangle.    
\label{eq:D}
\end{align}
Consider two adjacent vertices $v^*, {v^*}' \in \dVf$ as shown in Fig.~\ref{fig:adj}. Equation~\eqref{eq:D} implies
\begin{subequations}
\begin{align}
&J_e e^{-i\gamma^*/2} \langle q|v^*\rangle = \langle \bar q|v^*\rangle, \\
&J_e e^{-i\gamma^*/2} \langle q'|{v^*}'\rangle = \langle \bar q'|{v^*}'\rangle.
\end{align}
\label{eq:D1}
\end{subequations}
On the other hand, the definition of $D$ gives
\begin{subequations}
\begin{align}
   \langle {v^*}'| D | v^*\rangle  = e^{i\gamma/2} \langle\bar q' |  {v^*}' \rangle^*  \langle q |v^*\rangle,\label{eq:DD} \\
   \langle v^* | D | {v^*}'\rangle  =  e^{i\gamma/2} \langle \bar q |{v^*} \rangle^* \langle q' |{v^*}'\rangle. 
\end{align}
\end{subequations}
Substituting Eq.~\eqref{eq:D1}, we obtain
\begin{align}
    &\langle {v^*}'| D | v^*\rangle  = J_e e^{i(\gamma+\gamma^*)/2} \langle q' |  {v^*}' \rangle^*  \langle q |v^*\rangle, \notag \\
    &\langle v^* | D | {v^*}'\rangle^*  
     = J_e e^{-i(\gamma+\gamma^*)/2}  \langle q' |{v^*}'\rangle^* \langle q|{v^*}\rangle. \notag
\end{align}
Using the fact that $\gamma + \gamma^* = \pi$, the anti-Hermitian of $D|_{\ker(I - {D^*}^\dag)}$ follows directly.

\section{Proof of the frustration circulation law}\label{sec3}

In this section, we prove the frustration circulation law. To do this, we enlarge the space from the kernel of $I - {D^*}^\dag$, which is spanned by the frustrated dual vertices $\dVf$, to the space spanned by all dual vertices $\dV$. To proceed, we need to assign a vector $|v^*\rangle$ to each non-frustrated $v^*$. We may select one eigenvector of
\begin{align}
{D^*}^\dag |v^*\rangle = \lambda_{v^*} |v^*\rangle,   
\label{eq:D2}
\end{align}
with corresponding eigenvalue $\lambda_{v^*}$. The choice is arbitrary, but it must satisfy $|\lambda_{v^*}| = 1$ and $(\lambda_{v^*})^{d_{v^*}} = -1$, which are independent of the specific choice.

We start with a simpler case. For each fundamental cycle $C_v$ surrounding the vertex $v$, we have
\begin{equation}
   \sum_{(v^*\to{v^*}')\in C_v} \arg \langle {v^*}'| D | v^*\rangle = \pi,
\end{equation}
which is a direct consequence of the definition of $D$ in Eq.~\eqref{eq:DD}. The phases arising from $\langle q |v^*\rangle$ cancel out over the cycle, and the total geometric phase $\gamma$ sums to $2\pi$. Note that this result follows directly from the definition of $D$, independent of the choice of basis $|v^*\rangle$ or the frustration values $W_{v^*}$.

\begin{figure}[htbp]
  \includegraphics[width=0.5\linewidth]{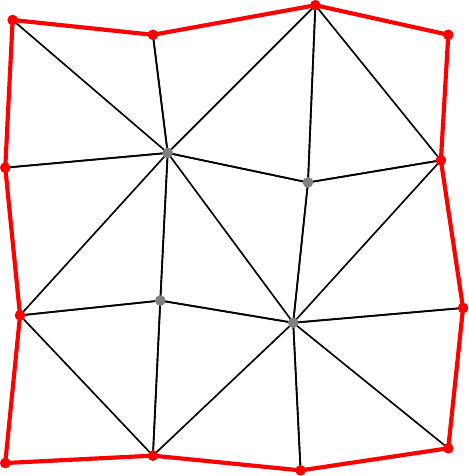}
  \caption{Illustration of a simple cycle $C$ (red).}
  \label{fig:cycle}
\end{figure}

We now consider a simple cycle $C$ over frustrated vertices, that is, all vertices on $C$ have $W = -1$ (see Fig.~\ref{fig:cycle}). The cycle $C$ can be decomposed into a linear combination of fundamental cycles as $C=\sum_v C_v$, leading to
\begin{align}
    \sum_{v\in\mathbf{F}(C)}&\sum_{(v^*\to{v^*}')\in C_v} \arg \langle {v^*}'| D | v^*\rangle = \sum_{(v^*\to{v^*}')\in C} \arg\langle {v^*}'| D | v^*\rangle \notag \\
    & + \sum_{(v^*,{v^*}') \in \E(C)} \arg \left(\langle {v^*}| D | {v^*}'\rangle\langle {v^*}'| D | v^*\rangle\right) \mod 2\pi,
\label{eq:C}
\end{align}
where $\mathbf{F}(C)$ and $\E(C)$ are the sets of faces and edges enclosed by $C$. Using Eq.~\eqref{eq:DD} and Eq.~\eqref{eq:D2}, we find
\begin{equation}
    \arg \left(\langle {v^*}| D | {v^*}'\rangle\langle {v^*}'| D | v^*\rangle\right) = \pi + \arg \lambda_{v^*} + \arg \lambda_{{v^*}'}\mod 2\pi, \notag
\end{equation}
which recovers the anti-Hermitian condition when $\lambda_{v^*} = \lambda_{{v^*}'} = 1$. Thus,
\begin{align}
    &\sum_{(v^*,{v^*}') \in \E(C)} \arg \left(\langle {v^*}| D | {v^*}'\rangle\langle {v^*}'| D | v^*\rangle\right) \notag\\
    &= |\E(C)| \pi + \sum_{v^* \in \dV(C)} d_{v^*} \arg \lambda_{v^*} \mod 2\pi, \notag \\
    & = (|\E(C)|  + n_{nf}(C)) \pi \mod 2\pi,
\label{eq:EC}
\end{align}
where $\dV(C)$ is the set of vertices enclosed by $C$, and $n_{nf}(C)$ is the number of non-frustrated vertices enclosed by $C$. Combining Eq.~\eqref{eq:C} and Eq.~\eqref{eq:EC}, we obtain
\begin{align}
    \sum_{(v^*\to{v^*}')\in C} \arg\langle {v^*}'| D | v^*\rangle  = (|\mathbf{F}(C)| - |\E(C)|  + n_{nf}(C)) \pi \mod 2\pi.
    \notag
\end{align}
Since $C$ is simply connected, we have
\begin{equation}
     n_{nf}(C) + n_f(C)  -|\E(C)| + |\mathbf{F}(C)| = 1,
     \notag
\end{equation}
where we use the fact that the number of edges and vertices on $C$ are equal and cancel each other. Therefore,
\begin{align}
    \sum_{(v^*\to{v^*}')\in C} \arg\langle {v^*}'| D | v^*\rangle  = (1 + n_{f}(C)) \pi \mod 2\pi.
    \notag
\end{align}

\section{Numerical Results}\label{sec4}
In this section, we present numerical results for a $100\times 100$ square lattice with periodic boundary condition and $p = 1/2$, averaged over $10^4$ realizations. Table~\ref{tab:num} shows the energy and entropy as the level increases, up to level-3. We find that the energy increases slightly with each level but remains relatively stable beyond level-1. Notably, our level-3 energy is higher than the values reported using the matching algorithm. This discrepancy may arise from the fact that matching-based methods typically impose a spatial cutoff, iterating until the energy converges numerically. In contrast, the level-3 corrections in our approach are highly non-local and likely capture long-range correlations beyond such cutoffs. 

For the entropy, we observe a systematic decrease with each added level, yielding values smaller than those reported in prior studies. It remains an open question whether, in the thermodynamic limit, the entropy density vanishes or converges to a finite non-zero value.

\begin{table}[htb!]
\caption{Energy density $e_0$ and entropy density $s_0$ truncated at each level.}
\begin{ruledtabular}
\begin{tabular}{lcccc}
 & Level 0 & Level 1 & Level 2 & Level 3 \\
\hline
$\overline{e_0}$ & $-3/2$ & $-1.418$ & $-1.407$ & $-1.396$ \\
$\overline{s_0}$ & $\frac{\ln2}{2}$ & $0.102$ & $0.066$ & $0.062$  \\
\end{tabular}
\end{ruledtabular}
\label{tab:num}
\end{table}

\begin{figure}[htbp]
  \includegraphics[width=1\linewidth]{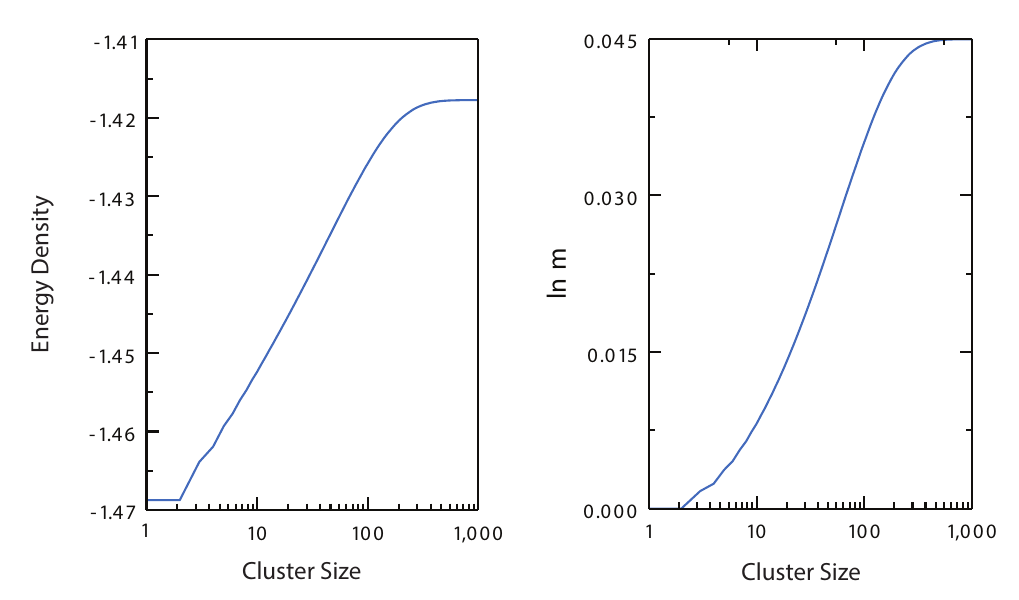}
  \caption{The cumulative summation of energy density (left) and $\ln \mathsf{m}$ as functions of cluster size.}
  \label{fig:cluster}
\end{figure}

Figure~\ref{fig:cluster} plots the cumulative energy density $e$ and $\ln \mathsf{m}$ as functions of cluster size at level-1, showing that convergence of the cluster expansion requires cluster sizes on the order of $100$. Similar plots for larger lattices indicate that this behavior is not a finite-size effect.

\bibliography{ref}

\end{document}


\title{Supplementary Materials}
\date{}
\author{}

\maketitle

\section{Ground state}
We now turn to the implications of our new formula for the RBIM. For the sake of technical convenience, our discussion will primarily focus on $\pm J$ disorder on a square lattice, i.e., $P(\tau) = (1-p)\delta(\tau-1) + p \delta(\tau+1)$. A straightforward generalization applies to arbitrary disorder. As we demonstrated earlier, the free energy of $\pm J$ bond disorder is dictated by the spectrum of Eq.~(\ref{eq:RBIM}), where only the disorder operator $\tilde d_*$ accounts for the bond disorders $\tau_e$.

We now turn our attention to the zero-temperature limit $\beta \to \infty$ of Eq.~(\ref{eq:RBIM}). Consequently, in this scenario, disorder operators dominate the spectrum.

\begin{equation}
    \ln\det(I-U) = -2M \ln(1+x^2) + \ln\det(I-{D^*}^\dag + 2x (x {D^*}^\dag - D)),
\end{equation}

Direct computation yields
\begin{align}
    \det(I-D_{v^*}^\dag) = 1+W_{v^*},
\end{align}
where the frustration $W_{v^*} \equiv \prod_{e\in P(v^*)}\tau_e = \pm 1$ is defined as the product of edge disorders around the plaquette of the corresponding dual vertices $v^*$. It is clear that when the plaquette is frustrated, i.e., $W_{v^*} = -1$, the determinant acquires a correction from the order operator $d$ with an order at most $e^{-2\beta J}$. On average, there is a 
$q(d)$ chance of $W_{v^*} = -1$, where
\begin{equation}
    q(d) \ldef  \frac{1-(2p-1)^d}{2}
\end{equation}

which results in a lower bound of the ground state energy density:

Since $H_0$ corresponds to the unweighted curl operator, it is block diagonalized into $n$ blocks, each associated with a vertex $v$, and each block contributes a zero mode given by

Given $D_{v*}$, one of the eigenvalue is $W_{v^*}=\prod_{e\in E(v^*)} w_{e}$, we have 
\begin{equation}
    \lambda_0 = (- W_{v^*})^{1/d_{v^*}},
\end{equation}
with 
\begin{equation}
    |v \rangle\ldef \frac{1}{\sqrt{d_v}} \sum_{v' \in N(v)} e_{v'} w_{v'},
\notag\end{equation}
which forms an $N^*$-dimensional kernel $\ker({D^*})$. Here, $d_v$ denotes the degree of vertex $v$, $N(v)$ is the set of neighbors of $v$, and $e_{v}$ is the unit vector with one at $v$ and zero elsewhere. 

$w_{i} = \bar w_{i,i+1}^{\alpha_1} \bar w_{i+1,i+2}^{\alpha_2}\ldots \bar w_{i-1,i}^{\alpha_{d_{v^*}}}$, where $\bar w_e = e^{i\gamma_e/2}w_e$ and 
\begin{equation}
   \left(\alpha_1,\ldots \alpha_{d_{v^*}} \right)= \left\{ \frac{d_{v^*}-1}{d_{v^*}},\ldots,0  \right\}
\end{equation}
Note while $D_{v^*}$ is non-Hermitian, meaning the left and right eigenvectors are generally distinct, they coincide for this eigenvalue, as given above, provided $|w_e| = 1$. Thus, we disregard this distinction between left and right eigenvectors in the following discussion.

Denoting the complementary space of $V_0$ as $V_\perp$, the $2m$-dimensional Hilbert space decomposes as $V_q = \ker({D^*}) \oplus \text{im}({D^*})$. Using this representation, we obtain
\begin{align}
\det(A+B) \approx {\det}'(A)\,\det\left((B-B\,A^+B)|_{\ker A}\right), 
\end{align}
where ${\det}'(A) \ldef \det\left(A|_{\text{im } A}\right)$, and $A^+$ denotes the Moore–Penrose inverse of $A$, that  annihilates the kernel of  $A$ and acts as a traditional inverse of  $A$ the subspace orthogonal to the kernel.

We set $A = I-{D^*}$ and $B = 2x (x {D^*}^\dag - D)$,
\begin{align}
&\ln\det(I-{D^*}^\dag + 2x (x {D^*}^\dag - D)) \notag\\
&\approx  \ln{\det}'(I-{D^*}^\dag) +  N_0^* \ln(2x)+ \ln\det\left((x - D) - 2x(x{D^*}^\dag - D) (I-{D^*})^+  (x {D^*}^\dag - D) |_{\ker{D^*}}\right)\notag\\
&=  \ln{\det}'(I-{D^*}^\dag) + N_0^* \ln(2x) + \ln \det\left(- iH_0+x(I-H_1)\right),
\end{align}
where 
\begin{equation}
    iH_0 \ldef  D|_{\ker{D^*}},\; H_1 \ldef 2(x{D^*}^\dag - D) (I-{D^*})^+  (x {D^*}^\dag - D)|_{\ker{D^*}},
\end{equation}
and 
\begin{equation}
     \ln{\det}'(I-{D^*}^\dag) = (N^*-N^*_0)\ln 2+ \sum_{v^* \in V^*_0} \ln d_{v^*}.
\end{equation}

Here, $H_0$ is a Hermitian matrix, with matrix element
\begin{equation}
     \langle {v^*}'|H_0|v^*\rangle  = -i\langle {v^*}'|D|v^*\rangle. 
\end{equation}

we are interested in 
\begin{align}
    -\beta F &= \frac{1}{2}\left(N \ln 2 + M \ln \cosh(\beta J) + \ln \det(I-U)  \right) \notag \\
    & = \frac{1}{2}\left(N \ln 2 + M \ln \cosh(\beta J) -2M \ln(1+x^2) + (N^*-N^*_0)\ln 2+ \sum_{v^* \in V^*_0} \ln d_{v^*} + N_0^* \ln(2x) + \ln \det(x - iH_0-x H_1) \right) \notag \\
    &=  \frac{1}{2}\left((N+N^*-M) \ln 2 + 2M \beta J + \sum_{v^* \in V^*_0} \ln d_{v^*} + N_0^* \ln x + \ln \det(x - iH_0-x H_1) \right)  + O(x) \notag \\
    & = \frac{1}{2}\left(2M \beta J + \ln\det(x(I-H_1)-iH_0)+ N_0^* \ln(x) + 2\ln 2 +\sum_{v^* \in V^*_-} \ln d_{v^*}\right) + O(x)
\end{align}

suppose the specturm of $H_0$ is given by 
\begin{equation}
p_0\delta(\lambda) + \rho(\lambda),
\end{equation}
 we have 
\begin{equation}
   \overline{\ln \det(x - iH_0) } = \overline{N_0^*} \left(p_0\ln x + \int_{-1}^{1} \ln|\lambda | \rho(\lambda) d\lambda \right) + \ln \det (I-H_1|_{\mathrm{ker}(H_0)})
\end{equation}

\begin{equation}
   \overline{\sum_{v^* \in V^*_-} \ln d_{v^*}} = \sum_{v^*\in V^*} q(d_{v^*})\ln d_{v^*} = N^* \left\langle{q(d^{*})\ln d^{*}}\right\rangle
\end{equation}

we are interested in 
\begin{align}
    -\beta F &=  \frac{1}{2}(N \ln 2 + 2M \beta J - M \ln 2 + \ln \det(I-U_{00}) + \ln\det(I-U_{\perp\perp}))  + O(x) \notag \\
    & = \frac{1}{2}(2M \beta J + \ln\det(x-iL)+ N_0^* \ln(x) + (N + N^* -M)\ln 2 +\sum_{v^* \in V^*_-} \ln d_{v^*})
\end{align}
thus
\begin{equation}
    -\beta f = \frac{1}{2} ( \bar d \beta J + \bar q (1+p_0) (\bar d/2-1)\ln x + (\bar d/2-1) \left\langle{q(d^{*})\ln d^{*}}\right\rangle +  \bar q (1-p_0) (\bar d/2-1) \int_{-1}^{1} \ln|\lambda | \rho(\lambda) d\lambda )
\end{equation}
as $-\beta f = -\beta e + s$, we have 
\begin{equation}
    e/J = \frac{1}{2} (\bar q (1+p_0) (\bar d-2)-\bar d) \geq  \frac{1}{2} (\bar q (\bar d-2)-\bar d),
\end{equation}
and
\begin{equation}
    s =\frac{(\bar d-2)\bar q}{4}  \left(\overline{\ln d^{*}} +   \int_{-1}^{1} \ln|\lambda | \rho(\lambda) d\lambda +  p_0\int \ln(1-\lambda)\rho'(\lambda)d\lambda\right),
\end{equation}
where $\bar q \ldef \sum_{d^*} q(d^*) P_d^*(d^*)$, and $\overline{\ln d^{*}}  \ldef \sum_{d^*} \ln d^* \frac{q(d^*)}{\bar q}P_d^*(d^*)$

numerical  $e/J =  -1.4007 \pm 0.0085$ and $s = 0.0709 \pm 0.006$.
our results: 
$e/J = (1+0.167828)/2 - 2 = -1.416086$ and 
$s = (2\ln(2)-1.21115(1-0.167828))/4 = 0.0946$

\begin{equation}
    e/J \geq -2+ \frac{1-(2p-1)^4}{2}.
\end{equation}
This finding is consistent with the results in Refs~\cite{vannimenus1977theory,grinstein1979ising} obtained using geometric approaches. 

\section{expansion}
\begin{align}
&\int_{0}^{\infty}\ln(x+v)P(x)\,dx = \int_{0}^{\infty}\ln x\,P(x)\,dx \\
&\quad - P(0)\,v\ln v + \left[\int_{0}^{\infty}\frac{P(x)-P(0)}{x}\,dx\right]v\\[1mm]
&\quad - \frac{P(0)}{2}\,v^2\ln v + \frac{1}{2}\left[\int_{0}^{\infty}\frac{P(x)-P(0)-xP'(0)}{x^2}\,dx\right]v^2 \\
&\quad + \mathcal{O}(v^3).
\end{align}
If 
\[
P(x)\sim C\,x^{-\alpha}\quad (0\le\alpha<1)
\]
for small \(x\), then the small‑\(v\) expansion is
\begin{align}
\int_{0}^{\infty}\ln(x+v)P(x)\,dx &= \int_{0}^{\infty}\ln x\,P(x)\,dx
-\frac{C}{1-\alpha}\,v^{\,1-\alpha}\ln v \\
&\quad + \; A\, v^{\,1-\alpha} + \mathcal{O}(v),
\end{align}

where 
\[
C=\lim_{x\to0}x^\alpha P(x)
\]
and \(A\) is a constant determined by the subleading behavior of \(P(x)\) near \(x=0\) and by the analytic contributions from \(x\gg v\).

\section{determinant formula}
When \(B\) is “small” (in a perturbative sense), one can often show that the proposed formula
\[
\det(A+B) = \det'(A)\,\det\Bigl((B-B\,A^+B)|_{\ker A}\Bigr)
\]
holds up to the lowest nontrivial order in \(B\). Here’s a bit more detail:

1. **Decomposition of the Space:**  
   Decompose the vector space as
   \[
   V=\operatorname{im}(A)\oplus\ker(A)
   \]
   so that in a suitable basis, one writes
   \[
   A = \begin{pmatrix} A_1 & 0 \\ 0 & 0 \end{pmatrix}
   \]
   with \(A_1\) invertible. Similarly, write
   \[
   B = \begin{pmatrix} B_{11} & B_{10} \\ B_{01} & B_{00} \end{pmatrix}.
   \]

2. **Determinant Expansion:**  
   In this block basis the sum becomes
   \[
   A+B = \begin{pmatrix} A_1+B_{11} & B_{10} \\ B_{01} & B_{00} \end{pmatrix}.
   \]
   For small \(B\) one expands the determinant using block–matrix formulas. Note that the contribution of \(B_{11}\) is in general a first–order correction:
   \[
   \det(A_1+B_{11}) \approx \det(A_1)\Bigl(1+\operatorname{tr}(A_1^{-1}B_{11})\Bigr).
   \]
   Meanwhile, the factor coming from the kernel block is
   \[
   \det\Bigl(B_{00} - B_{01}(A_1+B_{11})^{-1}B_{10}\Bigr).
   \]
   If you expand \((A_1+B_{11})^{-1}\) in \(B\), then to lowest order
   \[
   (A_1+B_{11})^{-1} =  A_1^{-1} - A_1^{-1}B_{11}A_1^{-1} + A_1^{-1}B_{11}A_1^{-1}B_{11}A_1^{-1} + \ldots,
   \]
   and hence
   \[
   \det\Bigl(B_{00} - B_{01}A_1^{-1}B_{10}\Bigr)
   \]
   appears. Notice that the term \(B_{01}A_1^{-1}B_{10}\) is quadratic in \(B\).

3. **Matching the Two Sides:**  
   The “reduced determinant” \(\det'(A)\) is exactly \(\det(A_1)\). Thus, to lowest nontrivial order in \(B\) the determinant becomes
   \[
   \det(A+B) \approx \det(A_1)\,\det\Bigl(B_{00} - B_{01}A_1^{-1}B_{10}\Bigr)
   \]
   which is exactly what you obtain by taking
   \[
   (B-B A^+B + BA^+B A^+B + \ldots )|_{\ker A} = B_{00} - B_{01}A_1^{-1}B_{10}  + B_{01}A_1^{-1}B_{11}A_1^{-1}B_{10}+\ldots
   \]
   In other words, even if the off–diagonal block \(B_{11}\) is not identically zero, its contribution only appears at higher order (since \(\operatorname{tr}(A_1^{-1}B_{11})\) multiplies an already small determinant factor if the kernel is nontrivial).

When \(B\) is small, the coupling between the image of \(A\) and its kernel gives only higher–order corrections. Up to first (or lowest nontrivial) order in \(B\) the identity
\[
\det(A+B) \approx \det'(A)\,\det\Bigl((B-B\,A^+B)|_{\ker A}\Bigr)
\]
holds. This is why in many perturbative applications—such as when handling zero–modes in physical theories—one finds that the effective determinant splits in this way.

Thus, while the exact equality may require extra conditions (like \(B\) not mixing the two subspaces), when \(B\) is small the leading–order behavior is captured by the formula you stated.

\section{Hierachy}
\begin{align}
    \ln\det(A^{(n)} + x B^{(n)}(x)) &= \ln \det\left(A^{(n)}_1+xB^{(n)}_{11}(x)\right)+\ln\det\left(x B^{(n)}_{00}(x)- x^2 B^{(n)}_{01} \left(A^{(n)}_1+xB^{(n)}_{11}(x)\right)^{-1} B^{(n)}_{10} \right), \\
    &=\dim \ker A^{(n)} \ln x + \ln {\det}'A^{(n)}  + \ln\det(A^{(n+1)} + x B^{(n+1)}(x)) + O(x),
\end{align}
where 
\begin{equation}
    A^{(n+1)} \ldef B^{(n)}(0)|_{\ker A^{(n)} },\;  B^{(n+1)}(x) \ldef \left.\frac{B^{(n)}(x)-B^{(n)}(0)}{x} - B^{(n)}(x) \left(\left.A^{(n)}+xB^{(n)}(x)\right|_{\mathrm{im}  A^{(n)}}\right)^{+} B^{(n)}(x)\right|_{\ker A^{(n)}},
\end{equation}
where 
\begin{equation}
    \left(\left.A^{(n)}+xB^{(n)}(x)\right|_{\mathrm{im}  A^{(n)}}\right)^{+} = A^{(n)+} - x  A^{(n)+} B^{(n)}(x) A^{(n)+} + \ldots
\end{equation}

\[
A^{(0)} = A,\; B^{(0)}(x)= - 2B + 2x (I - A),
\]

and 

\[
A^{(1)} = 2 \tilde A^{(1)} ,\; B^{(1)}(x)= 2 \tilde B^{(1)}(x),
\]
where
\[
\tilde A^{(1)} = -B|_{\ker A},\; B^{(1)}(x)= I - B(\tilde A^+ + x \tilde A^+ (B - x (I - A))\tilde A^++\ldots)B|_{\ker A},
\]
where $\tilde A^+ \ldef 2A^+$, or $\tilde A \ldef A/2$. By doing this, we have an additional term $\dim \ker A (\ln 2)$. consequently, we have

\[
\tilde A^{(2)} = (I-B \tilde A^+ B)|_{\ker \tilde A^{(1)}},\; B^{(2)}(x)=  -  B \tilde A^+ B \tilde A^+ B - (I-B \tilde A^+ B)|_{\ker A} \tilde A^{(1)+}(I-\tilde BA^+ B)|_{\ker A}|_{\ker \tilde A^{(1)}}  + O(x)
\]

\[
\tilde A^{(3)} = -  B \tilde A^+ B \tilde A^+ B - (I-B \tilde A^+ B)|_{\ker A} \tilde A^{(1)+}(I-\tilde BA^+ B)|_{\ker A}|_{\ker \tilde A^{(2)}},\; \ldots 
\]
also aware that $(I-\tilde BA^+ B)|_{\ker A}|_{\ker \tilde A^{(2)}}$ = 0, we have 
\[
\tilde A^{(3)} = -  B \tilde A^+ B \tilde A^+ B|_{\ker \tilde A^{(2)}},\; \ldots 
\]

since all these are anti-Hermitian, we have 
\begin{equation}
iH_n\ldef \tilde A^{(n)},
\end{equation}
we have
\begin{equation}
    H_1 = iD|_{\ker(I-{D^*}^\dag)}, H_2 = i (D \tilde A^+ D -I)|_{\ker(H_1)}
\end{equation}

\section{Moore--Penrose Pseudoinverse of a Cyclic Matrix}

\subsection{$(D_{v^*}^\dag)^{d_{v^*}}= -I$}
Consider the $d_{v^*} \times d_{v^*}$ matrix $D_{v^*}^\dag$, with $(D_{v^*}^\dag)^{d_{v^*}}= -I$. 

The Moore--Penrose pseudoinverse $(I-D_{v^*}^\dag)^+$ is same as the regular inverse
\begin{equation}
  (I-D_{v^*}^\dag)^+ = (I-D_{v^*}^\dag)^{-1} = \frac{1}{2}\sum_{n=0}^{d_{v^*}-1} (D_{v^*}^\dag)^n,
\end{equation}

\section{$(D_{v^*}^\dag)^{d_{v^*}}= I$}
Consider the $d_{v^*} \times d_{v^*}$ matrix $D_{v^*}^\dag$, with $(D_{v^*}^\dag)^{d_{v^*}}= I$. The eigenvalues of $D_{v^*}^\dag$ are the $n$th roots of unity:
\begin{equation}
\lambda_k = e^{2\pi i k/{d_{v^*}}}, \quad k=0,1,\dots,n-1.
\end{equation}
Thus, the eigenvalues of $A = I - D_{v^*}^\dag$ are
\begin{equation}
1 - \lambda_k.
\end{equation}
Because $\lambda_0 = 1$, we have a zero eigenvalue.

The Moore--Penrose pseudoinverse $(I-D_{v^*}^\dag)^+$ can be expressed as a polynomial in $D_{v^*}^\dag$:
\begin{equation}
(I-D_{v^*}^\dag)^+ = p(D_{v^*}^\dag) = \sum_{n=0}^{d_{v^*}-1} c_n (D_{v^*}^\dag)^n,
\end{equation}
with the conditions that for each eigenvalue 
\begin{equation}
    p(\lambda_k) = \begin{cases}
0, & \lambda = 1, \\[2mm]
\frac{1}{1-\lambda_k}, & \lambda_k \neq 1.
\end{cases}
\end{equation}
Using the discrete Fourier transform (i.e. Lagrange interpolation on the $n$-th roots of unity) one finds that the coefficients are given by
\begin{equation}
c_n = \frac{1}{d_{v^*}} \sum_{k=1}^{d_{v^*}-1} \frac{e^{-2\pi i nk/{d_{v^*}}}}{1-e^{2\pi i k/{d_{v^*}}}}
\end{equation}
or
\begin{equation}
c_n = \frac{d_{v^*}-1 - 2n}{2d_{v^*}}, \quad n = 0, 1, \dots, d_{v^*}-1.
\end{equation}
Thus, the final formula for the pseudoinverse is
\begin{equation}
(I-D_{v^*}^\dag)^+ = \frac{1}{2}\sum_{n=0}^{d_{v^*}-1} \frac{d_{v^*}-1 - 2n}{d_{v^*}} (D_{v^*}^\dag)^n,
\end{equation}

\section{alternate way of expansion}

Let $A \ldef I-{D^*}^\dag$ and $B \ldef D$, we have
\[
M(x) = A - 2x B + 2x^2 (I - A),
\]
and suppose \( \ker A \) has dimension \( p \), so we decompose the space as
\[
\mathbb{R}^n = \ker A \oplus (\ker A)^\perp,
\]
with corresponding block decompositions:
\[
A = \begin{pmatrix}
0 & 0 \\
0 & A_{11}
\end{pmatrix}, \quad
B = \begin{pmatrix}
B_{00} & B_{01} \\
B_{10} & B_{11}
\end{pmatrix}.
\]

Then the block structure of \( M(x) \) is:
\[
M(x) = \begin{pmatrix}
M_{00}(x) & M_{01}(x) \\
M_{10}(x) & M_{11}(x)
\end{pmatrix},
\]
where
\[
\begin{aligned}
M_{00}(x) &= -2x B_{00} + 2x^2 I = 2x(x I - B_{00}), \\
M_{01}(x) &= -2x B_{01}, \\
M_{10}(x) &= -2x B_{10}, \\
M_{11}(x) &= A_{11} - 2x B_{11} + 2x^2 (I - A_{11}).
\end{aligned}
\]

We compute the determinant using the formula
\[
\det M(x) = \det M_{00}(x) \cdot \det \left( M_{11}(x) - M_{10}(x) M_{00}(x)^{-1} M_{01}(x) \right).
\]

Let \( q = \dim \ker(B_{00}) \), and let \( P \) be the orthogonal projector onto \( \ker(B_{00}) \), \( Q = I - P \). Then:
\[
(xI - B_{00})^{-1} = \frac{1}{x} P - Q B_{00}^{-1} + \mathcal{O}(x).
\]
Thus,
\[
M_{00}(x)^{-1} = \frac{1}{2x^2} P - \frac{1}{2x} Q + \mathcal{O}(1).
\]

Now compute the Schur complement:
\[
\begin{aligned}
M_{10}(x) M_{00}(x)^{-1} M_{01}(x)
&= (-2x B_{10}) \left( \frac{1}{2x^2} P - \frac{1}{2x} Q \right) (-2x B_{01}) \\
&= 2 B_{10} P B_{01} + \mathcal{O}(x).
\end{aligned}
\]

Therefore,
\[
\begin{aligned}
K(x) &= M_{11}(x) - M_{10}(x) M_{00}(x)^{-1} M_{01}(x) \\
&= A_{11} - 2x B_{11} + 2x^2 (I - A_{11}) - 2 B_{10} P B_{01} + \mathcal{O}(x).
\end{aligned}
\]

Define the effective operator
\[
\tilde{A}_{11} = A_{11} - 2 B_{10} P B_{01},
\]
so that
\[
\ln \det K(x) = \ln \det \tilde{A}_{11} + \mathcal{O}(x).
\]

Now expand
\[
\det M_{00}(x) = (2x)^p \det(x I - B_{00}) = (2x)^p x^q \det{}'(B_{00}) (1 + \mathcal{O}(x)),
\]
where \( \det{}'(B_{00}) \) is the product of nonzero eigenvalues of \( B_{00} \). Taking logarithms gives:
\[
\ln \det M_{00}(x) = (p + q) \ln x + p \ln 2 + \ln \det{}'(B_{00}) + \mathcal{O}(x).
\]

Combining everything, we obtain the full expansion:
\[
\ln \det \left( A - 2x B + 2x^2 (I - A) \right)
= \ln \det \tilde{A}_{11} + \ln \det{}'(B_{00}) + p \ln 2 + (p + q) \ln x + \mathcal{O}(x).
\]

\section{equivlance of two expansion}
Let \( M(x) = A - 2x B + 2x^2 (I - A) \), and decompose the space as
\[
\mathbb{R}^n = \ker A \oplus (\ker A)^\perp,
\]
with \( \dim \ker A = p \). In this block decomposition:
\[
A = \begin{pmatrix} 0 & 0 \\ 0 & A_{11} \end{pmatrix}, \qquad
B = \begin{pmatrix} B_{00} & B_{01} \\ B_{10} & B_{11} \end{pmatrix}.
\]
Let \( q = \dim \ker(B_{00}) \), and let \( P \) be the orthogonal projector onto \( \ker(B_{00}) \), so that \( P = I \) on \( \ker(B_{00}) \) and \( B_{00} \) is invertible on \( Q = I - P \).

Consider the effective operator on \( (\ker A)^\perp \) that arises when factoring out \( M_{00} \):
\[
\widetilde{A}_{11} = A_{11} - 2 B_{10} P B_{01}.
\]
Alternatively, when factoring out \( M_{11} \), the Schur complement yields an operator acting on \( \ker(B_{00}) \):
\[
I - 2 B_{01} A_{11}^{-1} B_{10} \quad \text{(restricted to } \ker(B_{00}) \text{)}.
\]

We now use the matrix determinant lemma. Let
\[
A = A_{11}, \quad U = -2 B_{10} P, \quad V = B_{01}.
\]
Then
\[
A + U V = A_{11} - 2 B_{10} P B_{01} = \widetilde{A}_{11}.
\]
The matrix determinant lemma gives:
\[
\det(\widetilde{A}_{11}) = \det(A_{11}) \cdot \det\left( I + V A_{11}^{-1} U \right).
\]
Substituting in \( U, V \), we find:
\[
\det(\widetilde{A}_{11}) = \det(A_{11}) \cdot \det\left( I - 2 B_{01} A_{11}^{-1} B_{10} P \right).
\]
Since \( P \) is the identity on \( \ker(B_{00}) \), and the determinant is evaluated over this subspace, we may write:
\[
\det\left( I - 2 B_{01} A_{11}^{-1} B_{10} P \right)
=
\det\left( \left[ I - 2 B_{01} A_{11}^{-1} B_{10} \right] \big|_{\ker(B_{00})} \right).
\]
Therefore,
\[
\det(\widetilde{A}_{11}) = \det(A_{11}) \cdot \det\left( \left[ I - 2 B_{01} A_{11}^{-1} B_{10} \right] \big|_{\ker(B_{00})} \right).
\]

Taking logarithms, we find the equivalence:
\[
\ln \det(\widetilde{A}_{11}) = \ln \det(A_{11}) + \ln \det\left( \left[ I - 2 B_{01} A_{11}^{-1} B_{10} \right] \big|_{\ker(B_{00})} \right).
\]

This identity shows that whether the correction appears in \( \ker(B_{00}) \) as a factor
\[
\det\left( \left[ I - 2 B_{01} A_{11}^{-1} B_{10} \right] \big|_{\ker(B_{00})} \right),
\]
or as a shift of \( A_{11} \) to \( \widetilde{A}_{11} = A_{11} - 2 B_{10} P B_{01} \), the resulting contribution to \( \ln \det M(x) \) is the same. Hence, both block–factorization methods yield the same final expansion of \( \ln \det M(x) \) once all degeneracies are accounted for.

\section{Nested Hierachy}

Below is a more systematic, step-by-step proof of the ``nested-kernel'' (or ``nested Schur-complement'') formula for the matrix

\[
M(x) = A - 2x\,B + 2x^2 (I - A),
\]

under the assumption that each restricted operator remains singular on its kernel, so that the decomposition does not terminate.

We aim to prove the expansion:

\[
\ln \det M(x)
= \sum_{i=1}^\infty \dim \ker(O_i) \cdot \ln x
+ \sum_{i=1}^\infty \ln \det'\!(O_i),
\]

where each \( O_i = H_i|_{\ker(O_{i-1})} \) and the operators \( H_i \) are defined recursively by Schur complements, starting from

\[
H_1 = A, \quad H_2 = B, \quad H_3 = B - 2 B A^+ B,
\quad H_4 = (I - A) - 2 (I - A) (H_3^+) (I - A), \quad \text{etc.}
\]

Each \( H_i \) is the ``bare operator'' in the polynomial (either \( A \), \( B \), or \( I - A \)) minus a Schur-complement correction term, reflecting the partial inversion of the previous block.

We proceed step-by-step:

\textbf{Step 1: Decompose along \(\mathrm{Im}(A) \oplus \ker(A)\)}

We write \( V = \mathrm{Im}(A) \oplus \ker(A) \). In a basis adapted to this decomposition, the operators take block form:

\[
A = \begin{pmatrix}
\tilde{A} & 0 \\
0 & 0
\end{pmatrix}, \quad
B = \begin{pmatrix}
B_{11} & B_{12} \\
B_{21} & B_{22}
\end{pmatrix}, \quad
I - A = \begin{pmatrix}
I - \tilde{A} & 0 \\
0 & I
\end{pmatrix}.
\]

Then the full matrix becomes

\[
M(x) =
\begin{pmatrix}
\tilde{A} - 2x B_{11} + 2x^2(I - \tilde{A}) & -2x B_{12} \\
-2x B_{21} & -2x B_{22} + 2x^2 I
\end{pmatrix}.
\]

We now apply the Schur complement formula to compute \(\det M(x)\) block-by-block. Assuming \( \tilde{A} \) is invertible on its block, the determinant splits:

\[
\det M(x)
= \det(\tilde{A} - 2x B_{11} + \dots)
\cdot \det\left(-2x B_{22} + 2x^2 I - (-2x B_{21}) (\tilde{A}^{-1}) (-2x B_{12})\right).
\]

Simplifying the second factor:

\[
\det \left(2x^2 I - 2x B_{22} - 4x^2 B_{21} \tilde{A}^{-1} B_{12} \right)
= (2x)^{\dim \ker A} \cdot \det\left(B_{22} - 2 B_{21} A^+ B_{12} + \dots\right).
\]

This identifies the next effective operator on \( \ker(A) \) as

\[
H_3 = B - 2 B A^+ B.
\]

Thus, at this stage, we have:

\[
\ln \det M(x)
= \ln \det' A + \dim \ker A \cdot \ln(2x) + \ln \det'\left( B - 2 B A^+ B \right) + \cdots.
\]

\textbf{Step 2: Proceed to \(\ker(H_3)\)}

We now consider the space \( \ker(H_3) \subseteq \ker(A) \). On this subspace, the operator \( -2x B \) (with its correction) vanishes:

\[
H_3 v = \left( B - 2 B A^+ B \right) v = 0
\quad \Rightarrow \quad
-2x \left( B - 2 B A^+ B \right) v = 0.
\]

Hence, the next leading term on this subspace is:

\[
2x^2 (I - A),
\]

again with its own Schur complement correction, which gives the next effective operator:

\[
H_4 = (I - A) - 2 (I - A) (H_3^+) (I - A).
\]

\textbf{Step 3: Continue recursively}

Each step follows the same pattern:

\begin{itemize}
    \item Decompose the current subspace into \(\mathrm{Im}(H_i) \oplus \ker(H_i)\).
    \item Apply Schur complement to factor out the image block.
    \item The remainder acts on \(\ker(H_i)\), where the previous leading term vanishes.
    \item The new operator is of the form:
    \[
    \text{next bare term} - 2 \cdot \text{bare term} \cdot (H_i^+) \cdot \text{bare term}.
    \]
    \item The process continues with the next operator from the cycle: \(A \to B \to I - A \to B \to \cdots\).
\end{itemize}

Thus, we obtain the chain:

\[
\begin{aligned}
H_1 &= A, \\
H_2 &= B, \\
H_3 &= B - 2 B A^+ B, \\
H_4 &= (I - A) - 2 (I - A) H_3^+ (I - A), \\
H_5 &= B - 2 B H_4^+ B, \\
H_6 &= (I - A) - 2 (I - A) H_5^+ (I - A), \\
&\quad \vdots
\end{aligned}
\]

and define
\[
O_n = H_n \big|_{\ker(O_{n-1})}, \quad O_0 := 0, \quad \ker(O_0) := V.
\]

Each step contributes:

\[
\ln \det M(x)
= \sum_{n=1}^\infty \dim \ker(O_n) \cdot \ln x
+ \sum_{n=1}^\infty \ln \det'\!(O_n).
\]

\textbf{Why does \(-2x B + \text{correction}\) vanish on \(\ker(H_3)\)?}

We defined
\[
H_3 = B - 2 B A^+ B.
\]

Then for \(v \in \ker(H_3)\),
\[
H_3 v = 0 \quad \Rightarrow \quad B v = 2 B A^+ B v.
\]

Thus,
\[
\left[ -2x B + 4x B A^+ B \right] v = -2x (B - 2 B A^+ B) v = 0.
\]

So on \(\ker(H_3)\), the first-order part of \(M(x)\) vanishes, and the next active term is \(+2x^2 (I - A)\), which leads to the definition of \(H_4\).

\section{Bounding the Spectral Radius}

We consider the matrix \( B \) defined by
\[
B_{ij} = \frac{A_{ij} e^{i \phi_{ij}}}{\sqrt{d'_i d'_j}},
\]
where \( \phi_{ij} = -\phi_{ji} \) ensures that \( B \) is Hermitian. We assume that \( d'_i \geq d_i \), where \( d_i = \sum_j A_{ij} \) is the degree of node \( i \) in the underlying adjacency matrix \( A \). We aim to show that the spectral radius of \( B \) satisfies
\[
\rho(B) \leq 1.
\]

\subsection{Hermitian Property}
Since \( \phi_{ij} = -\phi_{ji} \), we verify that \( B \) is Hermitian:
\[
B_{ij}^* = \frac{A_{ij} e^{-i \phi_{ij}}}{\sqrt{d'_i d'_j}} = B_{ji}.
\]
Since a Hermitian matrix has real eigenvalues, it suffices to show that all eigenvalues lie within \([-1,1]\).

\subsection{Quadratic Form Representation}
For any nonzero vector \( x \in \mathbb{C}^n \), consider the quadratic form
\[
x^* B x = \sum_{i,j} \frac{A_{ij} e^{i \phi_{ij}}}{\sqrt{d'_i d'_j}} x_i^* x_j.
\]
Define a new vector \( y \) by
\[
y_i = \frac{x_i}{\sqrt{d'_i}},
\]
which rewrites the quadratic form as
\[
x^* B x = \sum_{i,j} A_{ij} e^{i \phi_{ij}} y_i^* y_j.
\]

\subsection{Bounding the Quadratic Form}
Using the triangle inequality and the fact that \(|e^{i \phi_{ij}}| = 1\), we get
\[
\left| \sum_{j} A_{ij} e^{i \phi_{ij}} y_j \right| \leq \sum_{j} A_{ij} |y_j|.
\]
Applying the Cauchy--Schwarz inequality and the definition of \( d_i \), we obtain
\[
\left| \sum_{i,j} A_{ij} e^{i \phi_{ij}} y_i^* y_j \right| \leq \sum_i d_i |y_i|^2.
\]
Since \( d'_i \geq d_i \), it follows that
\[
\sum_i d_i |y_i|^2 \leq \sum_i d'_i |y_i|^2.
\]
Noting that
\[
\sum_i d'_i |y_i|^2 = \sum_i d'_i \frac{|x_i|^2}{d'_i} = \sum_i |x_i|^2 = \|x\|^2,
\]
we conclude that
\[
|x^* B x| \leq \|x\|^2.
\]

\subsection{Spectral Radius Bound}
The Rayleigh quotient for a Hermitian matrix \( B \) is given by
\[
R(x) = \frac{|x^* B x|}{\|x\|^2}.
\]
Since we established that \( |x^* B x| \leq \|x\|^2 \), it follows that
\[
R(x) \leq 1.
\]
Since the spectral radius of a Hermitian matrix is the maximum absolute value of its Rayleigh quotient, all eigenvalues \( \lambda \) satisfy
\[
|\lambda| \leq 1.
\]
Thus, we conclude that \( \rho(B) \leq 1 \), completing the proof.

\section{Computational Complexity Analysis of Eigenvalue-Based Percolation Approach}

\subsection{Scaling of $\sum_i S_i^3$ in 2D Percolation}

In critical percolation, the probability that a cluster has size $S$ follows a power-law distribution:

\begin{equation}
P(S) \sim S^{-\tau}
\end{equation}

where the exponent $\tau$ in two-dimensional percolation is:

\begin{equation}
\tau = \frac{187}{91} \approx 2.055.
\end{equation}

The largest clusters scale as:

\begin{equation}
S_{\max} \sim N^{d_f / d}
\end{equation}

where:
- $d_f = \frac{91}{48} \approx 1.896$ is the fractal dimension of percolation clusters,
- $d = 2$ is the spatial dimension.

Thus, the largest cluster scales as:

\begin{equation}
S_{\max} \sim N^{0.948}.
\end{equation}

\section{Computing $\sum_i S_i^3$}

The total sum of cubic cluster sizes can be estimated as:

\begin{equation}
\sum_i S_i^3 \approx \int_{S_{\min}}^{S_{\max}} S^3 n(S) dS.
\end{equation}

Since the cluster number density follows:

\begin{equation}
n(S) \sim S^{-\tau},
\end{equation}

we substitute:

\begin{equation}
\sum_i S_i^3 \sim \int_{S_{\min}}^{S_{\max}} S^{3-\tau} dS.
\end{equation}

Using $\tau \approx 2.055$, we obtain:

\begin{equation}
\sum_i S_i^3 \sim \int_{S_{\min}}^{S_{\max}} S^{0.945} dS.
\end{equation}

Evaluating the integral:

\begin{equation}
\sum_i S_i^3 \sim S_{\max}^{1.945} \sim N^{0.945 \times 1.945} \approx N^{1.84}.
\end{equation}

\subsection{Alternative Derivation Using Fractal Exponents}

The scaling can also be derived using the relation:

\begin{equation}
d_f \times \frac{(4 - \tau)}{2}.
\end{equation}

Substituting known values:

\begin{equation}
d_f = \frac{91}{48}, \quad \tau = \frac{187}{91},
\end{equation}

we compute:

\begin{equation}
4 - \tau = 4 - \frac{187}{91} = \frac{364}{91} - \frac{187}{91} = \frac{177}{91}.
\end{equation}

Thus,

\begin{equation}
d_f \times (4 - \tau) = \frac{91}{48} \times \frac{177}{91} = \frac{177}{48}.
\end{equation}

Dividing by 2:

\begin{equation}
\frac{177}{96} \approx 1.84.
\end{equation}

\section{Frustration}

Frustration plays a central role in the physics of disordered magnetic systems, particularly in spin glasses, where local constraints prevent all interactions from being simultaneously satisfied. In the canonical $\pm J$ Ising spin glass model~\cite{edwards1975theory}, disorder is introduced via randomly distributed ferromagnetic and antiferromagnetic bonds. A fundamental quantity in understanding the energy landscape of such systems is the plaquette frustration variable
\begin{equation}
W_p = \prod_{\langle ij \rangle \in \partial p} \text{sign}(J_{ij}),
\end{equation}
which characterizes the $\mathbb{Z}_2$ flux threading each elementary plaquette~\cite{toulouse1977theory, wegner1971duality}. When $W_p = -1$, the plaquette is said to be \emph{frustrated}, indicating that no spin configuration can satisfy all bonds around that loop.

Previous studies have used the distribution of $W_p$ to classify disorder and to detect frustration networks~\cite{cieplak1990frustration, hartmann2001ground}. In particular, $W_p$ is invariant under local spin gauge transformations and plays the role of a Wilson loop in lattice gauge theory formulations~\cite{nishimori2001statistical, savit1980duality}. Algorithms for finding exact ground states in two-dimensional spin glasses, such as those based on minimum-weight perfect matching~\cite{bieche1980ground, barahona1982computational}, use $W_p = -1$ plaquettes to identify defects that must be paired by domain walls. However, $W_p$ has primarily been used as an input to combinatorial algorithms or to characterize disorder statistically, rather than as a structural object capable of directly determining thermodynamic quantities.

In this work, we demonstrate that the full ground state energy and residual entropy of the 2D $\pm J$ Ising model can be determined from the percolation geometry of frustrated plaquettes and the spectral properties of a \emph{magnetic adjacency matrix}. Specifically, we construct a matrix $\tilde{A}$ whose entries are
\begin{equation}
\tilde{A}_{ij} = e^{i \phi(i,j)} A_{ij}, \quad \phi(i,j) = -\phi(j,i),
\end{equation}
where $A_{ij}$ is the adjacency matrix restricted to the subgraph induced by the percolating clusters of frustrated plaquettes ($W_p = -1$), and $\phi(i,j)$ encodes the effective gauge field generated by the disorder configuration. We show that the number of zero modes of $\tilde{A}$ on each cluster scales linearly with cluster size, and that the total nullity across all such clusters determines both the ground state energy and entropy in the thermodynamic limit.

This approach bridges frustration geometry, spectral graph theory, and lattice gauge theory, and offers a new analytic window into spin glass ground states in two dimensions.

\section{Frustration, Holonomy, and Zero Modes in Gauged Graphs}

In this section, we explore the interplay between frustration, holonomy, and the nullity of magnetic adjacency matrices on graphs. We connect concepts from spin systems, supersymmetric quantum graphs, and spectral theory, culminating in a critical assessment of whether the number of frustrated plaquettes in planar graphs bounds the dimension of the kernel.

\subsection{Magnetic Adjacency Matrices and Holonomy}

Given a graph \( G = (V, E) \), we define a magnetic adjacency matrix \( A^{(\theta)} \in \mathbb{C}^{n \times n} \) by
\[
A_{ij}^{(\theta)} = A_{ij} e^{i\theta_{ij}}, \quad \theta_{ij} = -\theta_{ji} \in \mathbb{R},
\]
incorporates U(1) holonomy via a skew-symmetric phase matrix \( \theta_{ij} \) as Peierls phase \cite{Peierls1933}, yielding a Hermitian matrix. Here \( A_{ij} \in \{0,1\} \) is the unweighted adjacency matrix. The holonomy around a cycle \( C \subset G \) is given by
\[
\Phi(C) = \sum_{(i,j) \in C} \theta_{ij} \mod 2\pi.
\]

Holonomy is gauge-invariant and serves as a proxy for geometric frustration: when \( \Phi(C) \ne 0 \), destructive interference may prevent simultaneous minimization of all local constraints.

\subsection{Destructive Interference and Zero Modes}

A zero mode \( \psi \in \mathbb{C}^n \) satisfies \( A^{(\theta)} \psi = 0 \). When holonomy is nontrivial (e.g., \( \Phi(C) = \pi \)), the interference of complex phases around a cycle may cancel outgoing amplitudes at each vertex, producing zero modes.

In a triangle graph with uniform holonomy \( \Phi = \pi \), we can explicitly construct such a zero mode:
\[
\psi = \begin{pmatrix}
1 \\
e^{2\pi i / 3} \\
e^{\pi i / 3}
\end{pmatrix}
\quad \text{yields} \quad A^{(\theta)} \psi = 0.
\]
For \( \Phi \ne 0, \pi \), perfect destructive interference fails, and the kernel is trivial.

\subsection{Supersymmetric Quantum Graphs}

Supersymmetric (SUSY) quantum mechanics on graphs provides a natural framework to study such zero modes. The exterior derivative \( D^{(\theta)} \) defines a supercharge, and the SUSY Hamiltonian is given by
\[
H = Q^\dagger Q + Q Q^\dagger = 
\begin{pmatrix}
(D^{(\theta)})^\dagger D^{(\theta)} & 0 \\
0 & D^{(\theta)} (D^{(\theta)})^\dagger
\end{pmatrix}.
\]
The zero modes of this Hamiltonian lie in the kernel of \( D^{(\theta)} \) and \( (D^{(\theta)})^\dagger \), and their number is controlled by the Witten index, a topological invariant. Holonomy affects whether such zero modes survive.

\subsection{Connection to Frustrated Spin Systems}

Frustrated spin systems, such as the 2D \( \pm J \) Ising model, can be mapped to gauged graph models. In particular, frustration around a plaquette corresponds to a \( \pi \)-flux. The partition function of the planar Ising model with disorder can be expressed as a Pfaffian:
\[
Z = \sqrt{\det K},
\]
where \( K \) is a Kasteleyn matrix whose signs encode frustration. An odd number of frustrated plaquettes implies \( \det K = 0 \), guaranteeing at least a two-dimensional kernel.

\subsection{Frustrated Tight-Binding Models and Flat Bands}

In tight-binding models on planar graphs, such as graphene-like or kagome lattices, applying a magnetic field introduces frustration by embedding U(1) phases (Peierls phases) on the hopping terms. The resulting Hamiltonian takes the form of a gauged adjacency matrix:
\[
H = A^{(\theta)}, \quad A^{(\theta)}_{ij} = A_{ij} e^{i\theta_{ij}}, \quad \theta_{ij} = -\theta_{ji},
\]
where \( \theta_{ij} \in \mathbb{R} \) encodes the magnetic flux along edge \( (i,j) \), and the total holonomy \( \Phi(C) = \sum_{(i,j) \in C} \theta_{ij} \) around a plaquette \( C \) corresponds to the net magnetic flux through it.

When the flux per plaquette is nonzero — especially at \( \Phi = \pi \) — the system becomes frustrated. This frustration leads to \emph{destructive interference} around closed loops, which can support zero-energy eigenstates of the Hamiltonian. In many such lattices, this manifests as a \emph{flat band} at zero energy.

\paragraph{Flat Bands from Frustrated Plaquettes.}

A flat band is a set of eigenstates all having the same energy (typically zero), with high degeneracy. In frustrated planar lattices, these flat bands often consist of \emph{compact localized states} (CLSs), which are eigenstates with support only on a small, finite number of sites. These states result from destructive interference — their amplitudes are arranged so that outgoing amplitudes from each site cancel exactly due to the phase structure imposed by the flux.

\paragraph{Kagome Lattice Example.}

In the kagome lattice with uniform \( \pi \)-flux per triangle, each triangular plaquette becomes maximally frustrated. One can explicitly construct localized zero-energy eigenstates supported on a single triangle, with amplitudes that destructively interfere along the surrounding bonds. These states are linearly independent \emph{unless they overlap}, and each contributes to the nullity of \( A^{(\theta)} \). A similar construction appears in the Lieb lattice, where square and cross-shaped CLSs appear.

\paragraph{Bound on the Number of Zero Modes.}

Let \( P_\pi \) denote the number of plaquettes with \( \pi \)-flux (maximally frustrated). Then, under the condition that each such plaquette supports an independent localized zero mode, we have the inequality:
\[
\dim \ker A^{(\theta)} \leq P_\pi.
\]
This inequality is often saturated when:

\begin{itemize}
  \item The lattice geometry allows non-overlapping support for CLSs,
  \item The flux configuration is uniform or translation-invariant,
  \item The zero modes are not linearly dependent due to global symmetries.
\end{itemize}

However, if the CLSs overlap or if the flux pattern is non-uniform, linear dependencies among zero modes can occur, lowering the nullity. Thus, while the number of frustrated plaquettes provides a natural upper bound, the actual number of zero modes depends on geometric and algebraic constraints.

\paragraph{Spectral Implications.}

These flat bands correspond to a macroscopic degeneracy at zero energy, implying nontrivial topological or symmetry-enforced spectral structure. In interacting systems (e.g., flat-band ferromagnetism), they can induce spontaneous symmetry breaking or highly degenerate ground states.

For a detailed analysis of flat bands arising from frustration and destructive interference, see \cite{Mielke1991,Bergman2008}.

\subsection{References and Literature}

The key theoretical developments supporting these insights include Pfaffian representations of spin systems \cite{Kasteleyn1963,Fisher1966,Barahona1982}, flat-band models and magnetic frustration \cite{Mielke1991,Bergman2008}, and supersymmetric graph Hamiltonians \cite{Deshpande2019,Cooper2001}. While none provide a universal bound, they rigorously tie frustration and holonomy to spectral degeneracy in specific graph classes.

\section{proof of Hermitian}

\begin{figure}
\centering
  \includegraphics[width=0.5\linewidth]{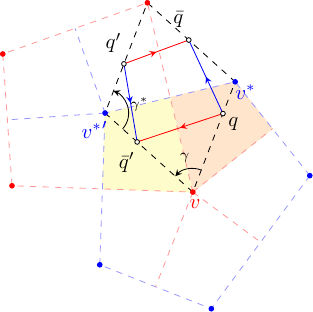}
  \caption{XXX}
  \label{fig:cross}
\end{figure}

where $L$ is a Hermitian matrix, defined as 
\begin{equation}
    i L_{{v^*}',v^*} = \langle {v^*}'|D|v^*\rangle 
\end{equation}

To prove the hermitian of $H_0$, we have  ${D^*}^\dag |v^*\rangle = |v^*\rangle$, or in terms of element
\begin{equation}
    \tau_e e^{-i\gamma^*/2} \langle q|v^*\rangle = \langle \bar q|v^*\rangle    
\end{equation}
\begin{equation}
   \tau_e  e^{-i\gamma^*/2} \langle q'|{v^*}'\rangle = \langle \bar q'|{v^*}'\rangle    
\end{equation}

basically we have 
\begin{align}
   \langle {v^*}'| D | v^*\rangle  = e^{i\gamma/2} \langle\bar q' |  {v^*}' \rangle^*  \langle q |v^*\rangle \\
   \langle v^* | D | {v^*}'\rangle  =  e^{i\gamma/2} \langle \bar q |{v^*} \rangle^* \langle q' |{v^*}'\rangle 
\end{align}
\begin{align}
    \langle {v^*}'| D | v^*\rangle  &=  e^{i\gamma/2} \langle\bar q' |  {v^*}' \rangle^*  \langle q |v^*\rangle \\
    & = \tau_e e^{i(\gamma+\gamma^*)/2} \langle q' |  {v^*}' \rangle^*  \langle q |v^*\rangle \\
    & = i \tau_e  \langle {v^*}'  |   q'\rangle  \langle q |v^*\rangle =  i \tau_e  \langle {v^*}'  |   \bar q'\rangle  \langle \bar q |v^*\rangle
\end{align}

\begin{align}
    \langle v^* | D | {v^*}'\rangle^*  &=  e^{-i\gamma/2}  \langle q' |{v^*}'\rangle^*  \langle \bar q |{v^*} \rangle \\
    & = \tau_e e^{-i(\gamma+\gamma^*)/2}  \langle q' |{v^*}'\rangle^* \langle q|{v^*}\rangle \\
    & = -i \tau_e  \langle {v^*}'  |   q'\rangle  \langle q |v^*\rangle = -i \tau_e  \langle {v^*}'  |   \bar q'\rangle  \langle \bar q |v^*\rangle
\end{align}

On the other hand,
\begin{equation}
      \langle {v^*}'|H_1|v^*\rangle  = 2\langle {v^*}'|(x {D^*}^\dag - D) (I-{D^*}^\dag)^+ (x{D^*}^\dag - D) |v^*\rangle = 2\langle {v^*}'|D (I-{D^*}^\dag)^+ D |v^*\rangle , 
\end{equation}
given $(I-{D^*}^\dag)^+ {D^*}^\dag |v^*\rangle = (I-{D^*}^\dag)^+ |v^*\rangle = 0$ and $\langle {v^*}'| {D^*}^\dag (I-{D^*}^\dag)^+ = \langle {v^*}'|  (I-{D^*}^\dag)^+  = 0$.
or 
\begin{equation}
    \langle {v^*}'|H_1|v^*\rangle  =  \sum_{n=0}^{d_{v^*}-1} \frac{d_{v^*}-1 - 2n}{d_{v^*}} \langle {v^*}'|D (D_{v^*}^\dag)^n D |v^*\rangle , 
\end{equation}

\section{Matching}

Let $G = (V, E)$ be a finite undirected graph with $n = |V|$ vertices and adjacency matrix $A$. A fundamental quantity associated with $G$ is its \emph{matching number} $\nu(G)$: the size of a maximum set of pairwise non-adjacent edges. The matching number relates deeply to the spectral structure of $A$, particularly to the dimension of its kernel, also called the \emph{nullity}.

For trees, a classical theorem of Cvetković \cite{cvetkovic1972algebraic} gives the exact relation:
\begin{equation}
    \dim \ker A = n - 2\nu(G),
\end{equation}
where $G$ is a forest (i.e., acyclic). In this case, nonzero eigenvalues appear in $\pm \lambda$ pairs due to bipartiteness, and unmatched vertices contribute to the kernel.

This paper explores whether this relation extends to graphs with cycles under the presence of \emph{magnetic frustration}.

Let $A^\phi \in \mathbb{C}^{n \times n}$ be a \emph{magnetic adjacency matrix}, defined on a graph $G = (V, E)$ with a phase function $\phi : E \to [-\pi, \pi]$ such that:
\[
A^\phi_{ij} =
\begin{cases}
e^{i\phi_{ij}} & \text{if } (i,j) \in E, \\
0 & \text{otherwise},
\end{cases}
\qquad \text{with } \phi_{ij} = -\phi_{ji}.
\]

This matrix is Hermitian: $(A^\phi)^\dagger = A^\phi$. The phase function defines a magnetic flux through each cycle in $G$:
\[
\Phi_C = \sum_{(i,j) \in C} \phi_{ij} \mod 2\pi.
\]

A cycle is called \emph{frustrated} if $\Phi_C \not\equiv 0 \mod 2\pi$. We define a graph as \emph{maximally frustrated} if $\Phi_C = \pi$ for every fundamental cycle $C$.

We propose the following conjecture based on extensive numerical evidence and analytic structure.

\paragraph{Conjecture.} Let $G$ be a connected graph with $n$ vertices and let $A^\phi$ be a magnetic adjacency matrix such that every cycle in $G$ has flux $\Phi_C = \pi$ (i.e., the system is maximally frustrated). Then the dimension of the kernel of $A^\phi$ is given by
\begin{equation}
    \boxed{
    \dim \ker A^\phi = n - 2\nu(G)},
\end{equation}
where $\nu(G)$ is the matching number of $G$.

This generalizes the tree case, where the result is known exactly. In our experiments, this formula holds precisely even for loopy graphs with complex frustration structure.

Let $M$ be a maximum matching of $G$, and decompose the vertex set into matched and unmatched parts: $V = V_M \sqcup V_0$ with $|V_M| = 2\nu$ and $|V_0| = n - 2\nu$. Under an appropriate permutation, the matrix $A^\phi$ can be block-decomposed:
\[
A^\phi =
\begin{pmatrix}
A_{MM} & A_{M0} \\
A_{0M} & A_{00}
\end{pmatrix}.
\]
The rank of $A_{MM}$ is generically $2\nu$, while the remaining structure (including frustrated cycles) contributes no additional rank due to interference cancellation.

The observation that magnetic frustration effectively suppresses rank suggests a deep connection between linear dependence, destructive interference, and topological constraints. Further work

\section{Existing result}

\begin{table}[h!]
\centering
\begin{tabular}{|l|l|l|l|l|}
\hline
Reference (Year) & Quantity & Value & Method & System Size (Lattice) \\
\hline
Jörg et al. (2011) \cite{Jorg_etal_2011} & Ground State Energy (per spin) & $-1.4015 \pm 0.0008\,J$ & Exact Algorithm & Up to $50\times 50$ \\
\hline
Nishimori \& Ortiz (N/A) \cite{Nishimori_Ortiz_ND} & Ground State Energy (per spin) & $\approx -1.40\,J$ & Numerical Estimate & N/A \\
\hline
Zhan et al. (2000) \cite{Zhan_etal_2000} & Ground State Energy (per spin) & $-1.4007 \pm 0.0085\,J$ & Flat Histogram Method & N/A \\
\hline
Jörg et al. (2011) \cite{Jorg_etal_2011} & Ground State Energy (per spin) & $-1.4065\,J$ & Exact Algorithm & $70\times 70$ \\
\hline
Lukic et al. (2018) \cite{Lukic_etal_2018} & Ground State Entropy Density (per spin) & $0.0714(2)$ & Exact Partition Function (Modular Arithmetic) & $\approx 100\times 100$ \\
\hline
Zhan et al. (2000) \cite{Zhan_etal_2000} & Ground State Entropy Density (per spin) & $0.0709 \pm 0.006$ & Flat Histogram Method & N/A \\
\hline
Hartmann (2001) \cite{Hartmann_2001} & Ground State Entropy Density (per spin) & $0.078(5)$ & Genetic Cluster-Exact Approximation & Up to $N=20^2$ \\
\hline
\end{tabular}
\caption{Comparison of ground state energy and entropy density values in 2D spin glass models.}
\label{tab:ground_state_properties}
\end{table}

\begin{table}[htb!]
\caption{\label{tab:num}Representative numerical results from previous studies across hierarchical levels.}
\begin{ruledtabular}
\begin{tabular}{lccccc}
 & Level 0 & Level 1 & Level 2 & Level 3 & Previous studies \\
\hline
$\overline{e_0}$ & $-3/2$ & $-1.418$ & $-1.407$ & $-1.396$ & $-1.4015(3)$ \\
$\overline{s_0}$ & $\frac{\ln2}{2}$ & $0.102$ & $0.066$ & $0.062$ & -- \\
\end{tabular}
\end{ruledtabular}
\end{table}

\bibliographystyle{unsrt}
\bibliography{ref}